\documentclass[aps,pra,twocolumn,amsmath,amssymb,showpacs]{revtex4-1}
\usepackage{graphicx}
\usepackage{amsmath}
\usepackage{bm}
\usepackage{cleveref}
\usepackage{xcolor}

\newcommand{\be}{\begin{equation}}  
\newcommand{\ee}{\end{equation}}
\newcommand{\ba}{\begin{array}}
\newcommand{\ea}{\end{array}}
\newcommand{\bea}{\begin{eqnarray}}
\newcommand{\eea}{\end{eqnarray}}

\newcommand{\bra}{\langle}
\newcommand{\ket}{\rangle}

\newcommand{\nn}{\nonumber}
\newcommand{\ca}{{\cal A}(\omega)}
\newcommand{\cad}{{\cal A}^{\dag}(\omega)}

\begin{document}

\title{Negative contributions to entropy production induced by quantum coherences
}

\author{C.L. Latune$^{1,2}$, I. Sinayskiy$^{1}$, F. Petruccione$^{1,2,3}$}
\affiliation{$^1$Quantum Research Group, School of Chemistry and Physics, University of
KwaZulu-Natal, Durban, KwaZulu-Natal, 4001, South Africa\\
$^2$National Institute for Theoretical Physics (NITheP), KwaZulu-Natal, 4001, South Africa\\
$^3$School of Electrical Engineering, KAIST, Daejeon, 34141, Republic of Korea}

\date{\today}
\begin{abstract}
The entropy production in dissipative processes is the essence of the arrow of time and the second law of thermodynamics. For dissipation of quantum systems, it was recently shown that the entropy production contains indeed two contributions: a classical one and a quantum one. Here we show that for degenerate (or near-degenerate) quantum systems there are additional quantum contributions which, remarkably, can become negative. Furthermore, such negative contributions are related to significant changes in the ongoing thermodynamics. This includes phenomena such as generation of coherences between degenerate energy levels (called horizontal coherences), alteration of energy exchanges and, last but not least, reversal of the natural convergence of the populations toward the thermal equilibrium state. Going further, we establish a complementarity relation between horizontal coherences and population convergence, particularly enlightening for understanding heat flow reversals. Conservation laws of the different types of coherences are derived. Some consequences for thermal machines and resource theory of coherence are suggested.

 \end{abstract}

\maketitle

\section{Introduction}

The study of entropy production is of paramount importance due to its intimate relation with the second law of thermodynamics \cite{Spohn_1978,Spohn_1978b,Alicki_1979}, 
 the emergence of irreversibility and the arrow of time in classical and quantum systems \cite{Parrondo_2009, Deffner_2011, Santos_2017, Brunelli_2018, Santos_2019, Batalhao_2018, Li_2019}. It is also related to other fundamental problems like reduction of performances in thermodynamic operations and thermal machines \cite{Barato_2015, Gingrich_2016, Pietzonka_2016, Pietzonka_2018, Guarnieri_2019, Timpanaro_2019, Su_2019, Holubec_2018}. 
As a recent remarkable development,  
for quantum systems undergoing dissipative processes (described either by Markovian evolutions or thermal operations), it was pointed out  \cite{Santos_2019,Francica_2019,Camati_2019} that the entropy production can be split in two contributions, an incoherent one (stemming from populations) and a coherent one (stemming from quantum coherences). 

Inspired by the above insight, by intriguing questions around reductions of entropy production in presence of  degenerate systems \cite{bathinducedcoh}, and by the special role of ``horizontal'' coherences (coherences between degenerate energy levels) in heat exchanges \cite{Savchenko_1998, Agarwal_2001, Dag_2016, Cuetara_2016, Cakmak_2017, Dag_2019, paperapptemp, bathinducedcohTLS, heatflowreversal}, we uncover additional quantum contributions to the entropy production. 
 We show that these extra quantum contributions, stemming from horizontal coherences and degenerate transitions, affect dramatically the ongoing thermodynamics.
 This includes the surprising possibility of reversing the natural convergence of the populations towards the equilibrium distribution. 
Furthermore, this phenomenon is associated to a negative contribution to the entropy production. 
Additionally, a second negative contribution can emerge, related to generation of horizontal coherences, which is indeed the underlying mechanism of well-known phenomena such as superradiance \cite{Skribanowitz_1973,Gross_1976, Raimond_1982, Gross_1982,Devoe_1996} and bath-induced entanglement (or dissipative generation of entanglement) \cite{Kim_2002, Plenio_2002, Benatti_2003, Muschik_2011, Krauter_2011,Cotlet_2014}, and affects heat exchanges. 
A complementary relation between horizontal coherences and reversal of the population's convergence is established: the consumption of one fuels the other, and reciprocally. It appears to be particularly insightful for heat flow reversals \cite{heatflowreversal}.   

The above results are firstly derived in the context of Markovian bath-driven dissipation. This viewpoint is extended by the end of the paper to include thermal operations \cite{Janzing_2000} (and even some a-thermal operations). Near degenerate systems are addressed in Appendix \ref{neardegenerate}. 
 Finally, whereas coherences between energy levels of different energy are globally conserved, horizontal coherences are not. Still, a conservation law for horizontal coherences together with population convergence can be established.

These negative contributions are contrasting from the always-positive quantum and classical contributions reported in \cite{Santos_2019,Francica_2019,Camati_2019}. 
Then, in addition to the above changes in the ongoing thermodynamics, one might expect further consequences in thermal machines and resource theory of coherence. \\ 

\section{Entropy production}\label{}
Throughout this paper we consider a system $S$ of degenerate Hamiltonian $H_S= \sum_n\sum_{i=1}^{l_n} e_n|n,i\ket\bra n,i|$ described by eigenenergies $e_n$, eigenstates $|n,i\ket$, and a degeneracy $l_n\geq 1$ for each energy level $n$. 
The system $S$ is assumed to undergo a dynamics described by a completely-positive-trace-preserving (CPTP) map \cite{Petruccione_Book, Jagadish_2018} $\Lambda_t$ such that at all time $t$ the density operator $\rho_t$ of $S$ is given by $\rho_t = \Lambda_t\rho_0$, where $\rho_0$ denotes the initial state of $S$. 
Furthermore, we consider that $\Lambda_t$ admits the thermal state $\rho^{\rm th}(\beta):=Z(\beta)^{-1}e^{-\beta H_S}$ as steady state (meaning that $\Lambda_t\rho^{\rm th} (\beta)=\rho^{\rm th} (\beta)$), where $Z(\beta):={\rm Tr} e^{-\beta H_S}$ is the partition function and $\beta$ is an inverse temperature. In the following, $\beta$ will correspond to the underlying inverse temperature of the system interacting with $S$. For now, keeping the discussion more general, we only require a CPTP map and a thermal steady state. 

\begin{figure}
\centering
\includegraphics[width=3.5cm, height=3.5cm]{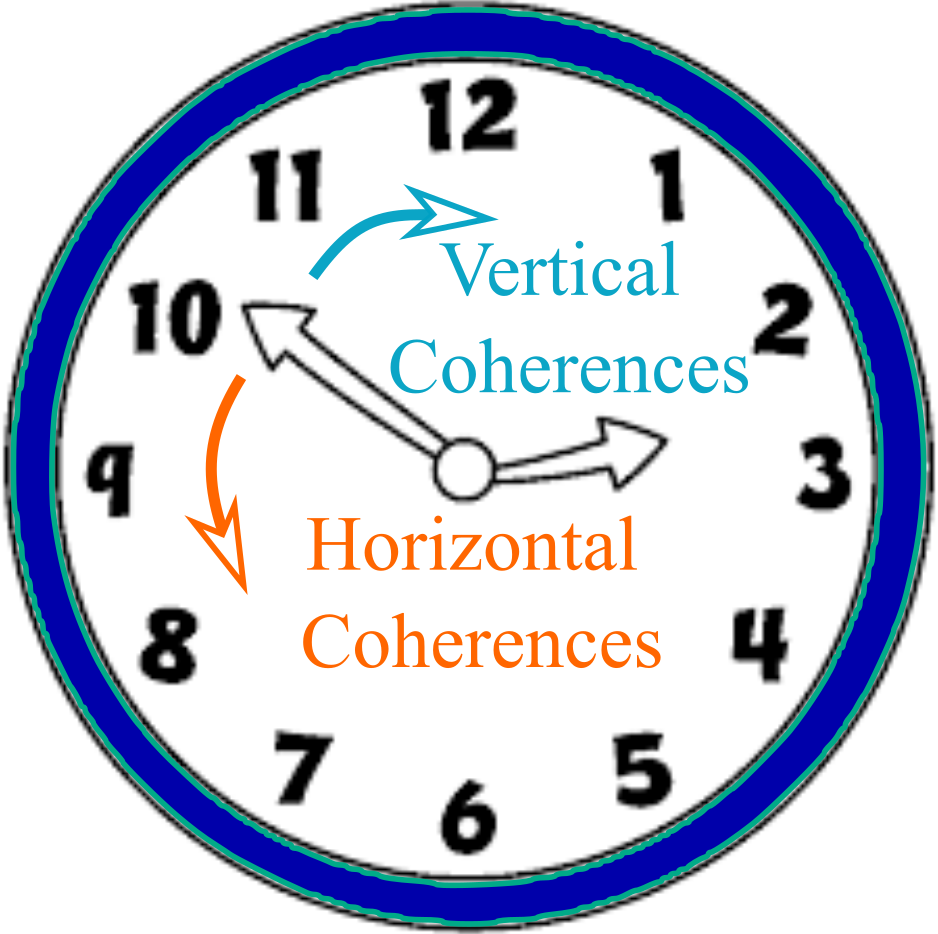}
\caption{{\bf Illustration representing the tendency of horizontal coherences (coherences between levels of same energy) to reverse the arrow of time.} 
The entropy production can be associated to the arrow of time \cite{Parrondo_2009, Deffner_2011, Santos_2017, Brunelli_2018, Santos_2019, Batalhao_2018, Li_2019}. Therefore, as seen throughout the paper, vertical coherences (coherences between levels of different energy) always strengthen the arrow of time (represented by the blue action), whereas horizontal coherences can tend to its reversal (represented by the orange action) due to their negative contributions to the entropy production.}
\label{reverse}
\end{figure}

The rate of change of the von Neumann entropy of $S$, defined by $S(\rho_t) :=-{\rm Tr} \rho_t\ln \rho_t$, can be split in two terms \cite{Spohn_1978,Spohn_1978b,Alicki_1979,Petruccione_Book,Kosloff_2013,Kosloff_2014},
\bea\label{srate}
\frac{d S(\rho_t)}{dt} &=& -{\rm Tr} \dot{\rho}_t \ln{ \rho_t }\nn\\
&=& -{\rm Tr}\dot{\rho}_t[\ln \rho_t-\ln \rho^{\rm th}(\beta)] - {\rm Tr}\dot{\rho}_t\ln \rho^{\rm th}(\beta),
\eea
where $\dot{\rho}_t$ is the time derivative of $\rho_t$. The first term in \eqref{srate} is identified as the rate of entropy production \cite{Spohn_1978,Spohn_1978b,Alicki_1979,Petruccione_Book,Kosloff_2013,Kosloff_2014}, 
\be\label{defentropyprod}
\Pi:=-{\rm Tr}\dot{\rho}_t[\ln \rho_t-\ln \rho^{\rm th}(\beta)],
\ee
 and is always positive (due to the contraction of the relative entropy under completely positive and trace preserving maps \cite{Lindblad_1975,Petruccione_Book}, see more details in the following). The second term in \eqref{srate}, of arbitrary sign, is the rate of entropy flow 
$\Phi :=- {\rm Tr}\dot{\rho}_t\ln \rho^{\rm th} (\beta)=\beta \dot E_S$ and is identified as heat exchanges, with the internal energy of $S$ defined as $E_{S}(t):={\rm Tr} \rho_{t}H_S$.\\

\section{The three contributions to entropy production} 
\begin{figure}
\centering
\includegraphics[width=6cm, height=6cm]{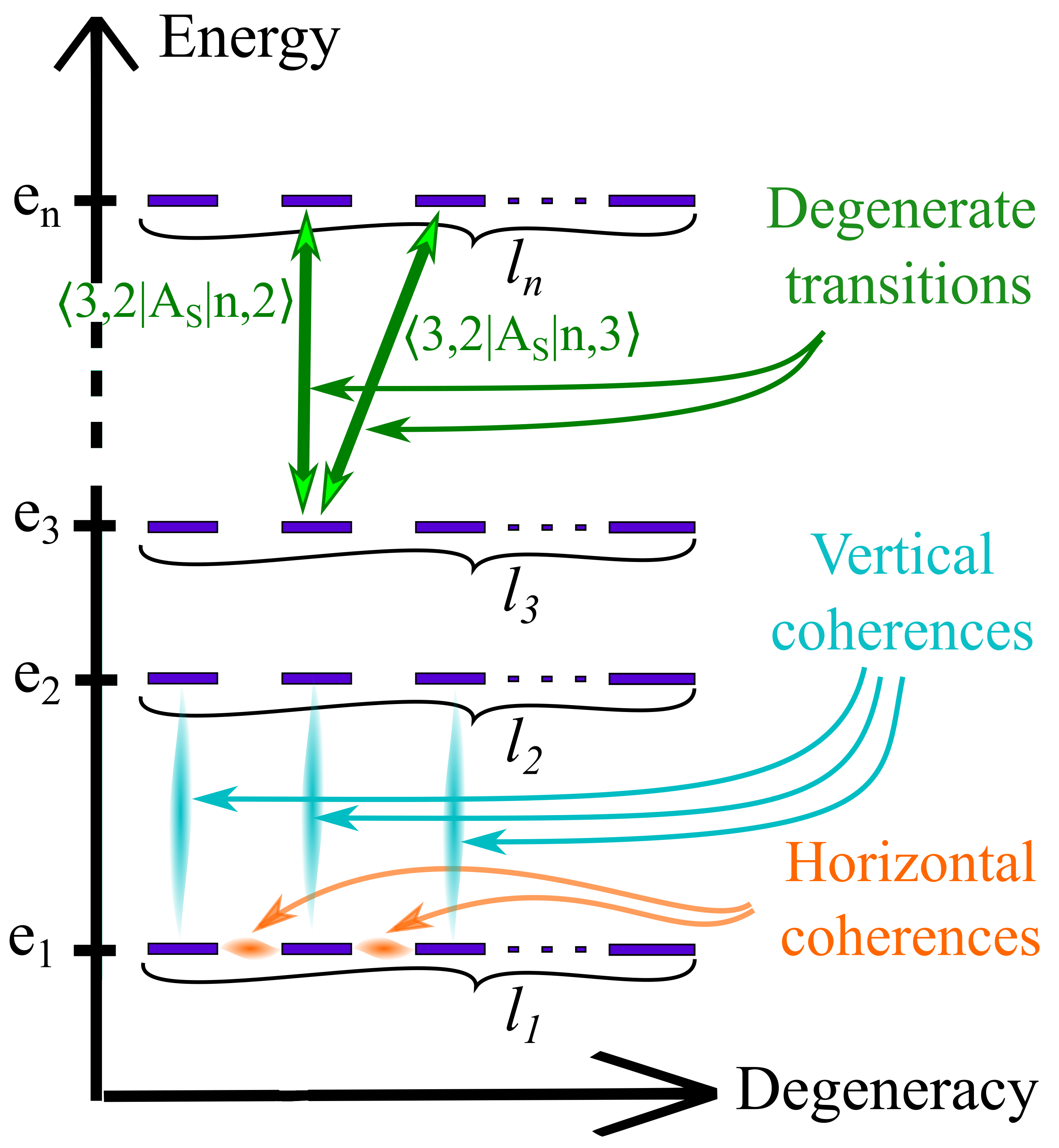}
\caption{{\bf Energy level structure of $S$}. Coherences between levels of same energy are called {\it horizontal} coherences whereas coherences between levels of different energy are called {\it vertical} coherences, in a direct reference to their graphical representation. The green double arrows represent degenerate transitions, playing a central role in the emergence of negative contributions to entropy production.}
\label{energylevels}
\end{figure}

We denote by $\rho_{t|_{\rm D}}$ the diagonal matrix obtained by canceling all non-diagonal elements of $\rho_t$ when written in the energy eigenbasis $|n,i\ket$. 
Then, we defined $\rho_{t|_{\rm BD}}$ as the block-diagonal matrix obtained by canceling only coherences between levels of different energies. In other words, $\rho_{t|_{\rm D}}:= \sum_n\sum_{i=1}^{l_n} \bra n,i|\rho_t|n,i\ket|n,i\ket\bra n,i|$ while $\rho_{t|_{\rm BD}}:=\sum_n \pi_n\rho_t\pi_n$, where $\pi_n:=\sum_{i=1}^{l_n}|n,i\ket\bra n,i|$. In the remainder of the paper, coherences between levels of same energy is referred to as {\it horizontal} coherences. By contrast, coherences between levels of different energies is called {\it vertical} coherences. This is in reference to the respective position of the different energy levels in an energy diagram (see Fig. \ref{energylevels}). Note that sometimes a different terminology is used (``energetic'' and ``non-energetic'' coherences), but here we prefer to introduce this new terminology to avoid possible confusion with the reference basis.  

With the above definitions we can decompose the entropy production as
\bea\label{decomp}
\Pi = - \dot{\cal C}_{\rm v} -\dot{\cal C}_{\rm h} -\dot{\cal D}_{\rm th}.
\eea
The first term in the above identity is the time derivative of the relative entropy of vertical coherence which we defined as ${\cal C}_{\rm v}(t):= S(\rho_{t|_{\rm BD}})-S(\rho_t)$ (equivalent to the definition in \cite{Baumgratz_2014} for non-degenerate systems). The quantity ${\cal C}_{\rm v}(t)$ is a measure of {\it vertical} coherences contained in $\rho_t$ \cite{Baumgratz_2014}. Note that this quantity was already introduced under the name of {\it relative entropy of asymmetry} \cite{Manzano_2019, Marvian_2016a, Marvian_2016b} in a context of resource theory. 
The second term in \eqref{decomp} is the time derivative of the relative entropy of horizontal coherence that we define as ${\cal C}_{\rm h}(t): =S(\rho_{t|_{\rm D}})-S(\rho_{t|_{\rm BD}})$. In analogy with the first term, the quantity ${\cal C}_{\rm h}(t)$ is a measure of {\it horizontal} coherences contained in $\rho_t$.  
The last term $-\dot {\cal D}_{\rm th}$ is the time derivative of $-{\cal D}_{\rm th} (t):= -S[\rho_{t|_{\rm D}}|\rho^{\rm th}(\beta_B)]$ defined through the relative entropy $S(\sigma|\rho):={\rm Tr} \sigma (\ln\sigma-\ln\rho)$ which establishes a measure of distance between any two density operators $\sigma$ and $\rho$ \cite{Nielsen_Book}. Therefore, $S[\rho_{t|_{\rm D}}|\rho^{\rm th}(\beta_B)]$ measures how far the population distribution is from the thermal equilibrium distribution, and $-\dot {\cal D}_{\rm th}$ is the rate -- or velocity -- to which the population distribution converge to the thermal equilibrium distribution. Moreover, defining by ${\cal F}_{\rm D}(t) := E_S(t) - S(\rho_{t|_{\rm D}})/\beta_B$ the {\it diagonal} -- or {\it classical} -- free energy, we have the interesting relation $\dot{\cal D}_{\rm th}=\beta_B\dot{\cal F}_{\rm D}$. 
One should keep in mind that degenerate Hamiltonians admit an infinite number of energy eigenbasis. While ${\cal C}_{\rm v}$ does not depend on the choice of the energy eigenbasis, ${\cal C}_{\rm h}$ and ${\cal D}_{\rm th}$ do. This is not an issue as long as one sticks to a given eigenbasis. 
Typically, one chooses the ``natural'' basis, representing localised excitations, as for instance the local basis of a many-body system.

Note that thanks to the identities \eqref{identity1} and \eqref{identity2} of Appendix \ref{apptimeder}, the relative entropy of vertical and horizontal coherences can be rewritten as ${\cal C}_{\rm v}(t) ={\rm Tr}\rho_t\left[\ln\rho_t-\ln\rho_{t|_{\rm BD}}\right] =S(\rho_t|\rho_{t|_{\rm BD}})$ and ${\cal C}_{\rm h}(t) ={\rm Tr} \rho_t\left[\ln\rho_{t|_{\rm BD}} -\ln \rho_{t|_{\rm D}}\right]=S(\rho_{t|_{\rm BD}}|\rho_{t|_{\rm D}})$. One can also verify the following identities (see Appendix \ref{apptimeder}),  
\bea\label{identities}
&&\dot{\cal C}_{\rm v}={\rm Tr}\dot\rho_t\left[\ln\rho_t-\ln\rho_{t|_{\rm BD}}\right],\nn\\
 &&\dot{\cal C}_{\rm h}={\rm Tr} \dot\rho_t\left[\ln\rho_{t|_{\rm BD}} -\ln \rho_{t|_{\rm D}}\right],\nn\\ 
&&\dot {\cal D}_{\rm th} = {\rm Tr} \dot\rho_t \left[\ln \rho_{t|_{\rm D}} - \ln \rho^{\rm th}(\beta)\right],
\eea
leading to \eqref{decomp}. \\

It is quite remarkable that the entropy production is related to the rate at which vertical coherences are consumed \cite{Santos_2019}, corresponding to the term $-\dot {\cal C}_{\rm v}$, but additionally to the rate at which horizontal coherences are consumed, expressed by the term $-\dot {\cal C}_{\rm h}$, topped by the velocity of the population convergence to the thermal equilibrium distribution, $-\dot {\cal D}_{\rm th}$.  

\section{Bath-driven dissipation} 
In the following we study the role and behaviour of each contribution for one of the most common situation in thermodynamics and quantum dynamics: interaction with a bath.
 We will see that the presence of horizontal coherences has interesting consequences.
We consider that the system $S$ is interacting with a stationary bath $B$ \cite{Alicki_2014,Alicki_2015,paperapptemp}, meaning that the bath state $\rho_B$ commute with its free Hamiltonian $H_B$, $[\rho_B,H_B]=0$. 
We assume that the system-bath coupling is of the form $V=g  A_{S}A_B$, 
where $g$ corresponds to the effective coupling strength, and $A_{S}$ and $A_B$ are observables of $S$ and $B$, respectively. More details regarding the physics behind of such coupling and its relation with an underlying notion of indistinguishability can be found in the Appendix \ref{appcolcoupling}. 
 %
 
Under weak coupling, the Born and Markov approximations are legitimate \cite{Cohen_Book,Petruccione_Book} so that one can derive the following master equation using the secular approximation \cite{Cohen_Book,Petruccione_Book} for the reduced dynamics of the system (in the interaction picture)
\bea\label{me}
\dot{\rho}_t &=& {\cal L} \rho_t \nn\\
&:=&  \sum_{\omega} \Gamma(\omega) \left[{\cal A}(\omega)\rho_t {\cal A}^{\dag}(\omega) -{\cal A}^{\dag}(\omega){\cal A}(\omega)\rho_t\right] + {\rm h.c.},\nn\\
\eea
where $\Gamma(\omega)=\int_0^{\infty}ds e^{i\omega s}{\rm Tr}\rho_B A_B(s)A_B$, and $A_B(s)$ is the bath operator $A_B$ in the interaction picture (with respect to the free Hamiltonian $H_B$). The jump operators ${\cal A}(\omega)$ are defined by \cite{Petruccione_Book} ${\cal A}(\omega)=\sum_{e_{n'}-e_n=\omega}  \pi_n A_S \pi_{n'}$.
One equilibrium state of the dynamics \eqref{me} is the thermal state $\rho^{\rm th}(\beta_B)$, where $\beta_B$ is the bath inverse temperature (or apparent temperature \cite{paperapptemp,autonomousmachines} for a non-thermal stationary state). 

Finally, one important characteristic of the physics described by \eqref{me} is the independence of the vertical coherences' dynamics from the populations and the horizontal coherences whereas the horizontal coherences' dynamics is coupled to the populations (see Appendix \ref{appcolcoupling}). This observation has deep implications as we will see in the following.\\

\subsection{Negative contribution of $-\dot {\cal C}_{\rm h}$} \label{secnegcontrib}
For non-degenerate systems $\rho_{t|_{\rm D}}=\rho_{t|_{\rm BD}}$, so that ${\cal C}_{\rm h}=0$ at all times. Then, $-\dot {\cal C}_{\rm v}$ and $-\dot {\cal D}_{\rm th}$ become equivalent to the coherent and diagonal contributions introduced in Eq. (12) of \cite{Santos_2019}. In Appendix \ref{appnondegen} we show a simple proof of their positivity, which is similar to the following one for $-\dot {\cal C}_{\rm v}$ in a context of degenerate systems. Namely,
\bea\label{mainposen}
-\dot {\cal C}_{\rm v} &=& -\frac{d}{dt} S(\rho_t|\rho_{t|_{\rm BD}})\nn\\
 &=&-\lim_{dt\rightarrow 0} \frac{1}{dt} \Big[S(\rho_{t+dt}|\rho_{t+dt|_{\rm BD}})-S(\rho_{t}|\rho_{t|_{\rm BD}})\Big]\nn\\
&=&-\lim_{dt\rightarrow 0} \frac{1}{dt} \Big[S(e^{dt {\cal L}}\rho_{t}|e^{dt{\cal L}}\rho_{t|_{\rm BD}})-S(\rho_{t}|\rho_{t|_{\rm BD}})\Big],\nn\\
\eea
which is always positive since the relative entropy is contractive under completely positive and trace preserving maps \cite{Lindblad_1975,Petruccione_Book} (the map generated by ${\cal L}$ defined in \eqref{me} being completely positive and trace preserving). 
 The crucial step in \eqref{mainposen} is 
\be\label{equalitybis}
 \rho_{t+dt|_{\rm BD}} = e^{dt {\cal L}}\rho_{t|_{\rm BD}}, 
 \ee
  holding since, as mentioned above, the dynamics of the populations and horizontal coherences (both contained in $\rho_{t|_{\rm BD}}$) are independent from the vertical coherences (see Appendix \ref{appcolcoupling}). 
However, the dynamics of the populations and the horizontal coherences are coupled, which implies 
\be
\rho_{t+dt|_{\rm D}} \ne e^{dt {\cal L}}\rho_{t|_{\rm D}}.
\ee
Then, 
\bea\label{cnen}
&-&\dot {\cal C}_{\rm h}  =-\lim_{dt\rightarrow 0} \frac{1}{dt} \Big[S(\rho_{t+dt|_{\rm BD}}|\rho_{t+dt|_{\rm D}})-S(\rho_{t|_{\rm BD}}|\rho_{t|_{\rm D}})\Big]\nn\\
&&\ne-\lim_{dt\rightarrow 0} \frac{1}{dt} \Big[S(e^{dt {\cal L}}\rho_{t|_{\rm BD}}|e^{dt{\cal L}}\rho_{t|_{\rm D}})-S(\rho_{t|_{\rm BD}}|\rho_{t|_{\rm D}})\Big],\nn\\
\eea
breaking down the guarantee of positivity of $-\dot {\cal C}_{\rm h}$.
 Then, the guarantee being broken, one can be sure the worst can happen: $-\dot {\cal C}_{\rm h}$ can become negative. An example of that in a quite general situation follows.

Considering the dynamics described by \eqref{me}, we assume for instance that the system $S$ is initially in a thermal state $\rho_0=\rho^{\rm th}(\beta_0)$. The state of $S$ at a later time is therefore given by (in the interaction picture)
\bea\label{expansion}
\rho_t &=& e^{t{\cal L}}\rho_0 \nn\\
&=& \rho_0 + t {\cal L} \rho_0 + {\cal O}(\Gamma^2t^2),
\eea
where $\Gamma := {\rm max}_{\omega} |\Gamma(\omega)|$ characterises the dissipation rate suffered by $S$. 
Then, for times much smaller than $\Gamma^{-1}$, the state of $S$ is well approximated by the first two terms of \eqref{expansion}. Using the fact that $S$ is initially in a thermal state at inverse temperature $\beta_0$ and the following identity $\ca\rho^{\rm th}(\beta_0)=e^{-\omega\beta_0}\rho^{\rm th}(\beta_0)\ca$, one obtains
\bea
\rho_t &=& \rho^{\rm th}(\beta_0)\Bigg\{ 1 + t \sum_{\omega>0}G(\omega)\big(e^{-\omega\beta_0}-e^{-\omega\beta_B}\big)\nn\\
&&\hspace{1.4cm}\times \big[\ca\cad-e^{\omega\beta_0}\cad\ca\big]\Bigg\} \nn\\
&&+ {\cal O}(\Gamma^2t^2),
\eea
where $G(\omega):= \Gamma(\omega) +\Gamma^{*}(\omega)$ is related to the inverse bath temperature (or apparent temperature \cite{paperapptemp,autonomousmachines}) $\beta_B$ through the relation $G(-\omega)/G(\omega) = e^{-\omega\beta_B}$.  
 Denoting by $\{|n,i\ket\}$ the chosen energy eigenbasis (for instance the ``natural'' basis), we will call {\it degenerate} a transition involving two degenerate levels $|n,i_1\ket$, $|n,i_2\ket$, and a third level $|n',i'\ket $ (see Fig. \ref{energylevels}), so that  $\bra n',i'|A_S|n,i_1\ket \ne 0$ and $\bra n',i'|A_S|n,i_2\ket \ne 0$. Then, in the presence of degenerate transitions, one can show that the terms $\ca \cad$ and $\cad\ca$ contain horizontal coherences. Indeed, for $\omega=e_{n'}-e_n$,
\bea
&&\bra n,i_1|\ca\cad|n,i_2\ket \nn\\
&&= \sum_{e_{m'}-e_m=\omega}\bra n,i_1|\pi_mA_S\Pi_{m'}A_S\pi_m|n,i_2\ket \nn\\
&&= \bra n,i_1|A_S\pi_{n'}A_S|n,i_2\ket \ne 0.
\eea
Similarly, we have also $\bra n,i_1|\cad\ca|n,i_2\ket \ne 0$. This implies $\bra n,i_1|\rho_t|n,i_2\ket \ne 0$ when $\beta_B\ne\beta_0$. In other words, the presence of degenerate transitions 
 {\it generates} horizontal coherences in $\rho_t$. This is the underlying common mechanism of bath-induced coherences in multi-level atoms \cite{Dodin_2016, Patnaik_1999, Koslov_2006, Tscherbul_2014}, crucial in superradiance \cite{Skribanowitz_1973,Gross_1976, Raimond_1982, Gross_1982,Devoe_1996} and bath-induced entanglement \cite{Kim_2002, Plenio_2002, Benatti_2003, Muschik_2011, Krauter_2011, Cotlet_2014}. 

 Therefore, since the presence of horizontal coherences implies $\rho_{t|_{\rm BD}} \ne \rho_{t|_{\rm D}}$, we have
\be
-{\cal C}_{\rm h}(t) = -S(\rho_{t|_{\rm BD}}|\rho_{t|_{\rm D}})<0,
\ee
leading to the negativity of $-\dot {\cal C}_{\rm h}$. In particular, a conclusion of this paragraph is that the phenomenon of bath-induced coherences known in the diverse contexts just mentioned are all associated with negative contribution to the entropy production.  
Based on identifications of negative entropy production as reversal of the arrow of time \cite{Partovi_2008, Jennings_2010, Micadei_2019, Henao_2018}, one can interpret the above result as a tendency of horizontal coherences to reverse the arrow of time, while vertical coherences always re-enforce it, see Fig. \ref{reverse}. Note that this is only a tendency since the total entropy production remains positive.\\

{\it Illustration: reduction of irreversibility}. 
Extending the conclusions of the previous paragraph, we show in the following an illustration where horizontal coherences reduce the irreversibility of dissipative processes.
 We consider an ensemble of $n$ spins of dimension $s$ indistinguishable from the point of view of the bath and therefore following a dynamics described by \eqref{me} (see also Appendix \ref{appcolcoupling}). The thermodynamic properties emerging from the resulting collective dissipation were studied in details in \cite{bathinducedcoh}. Using some results of \cite{bathinducedcoh} one can show (see Appendix \ref{appss}) that for an ensemble initially in a thermal state at inverse temperature $\beta_0$, the contribution to the entropy production from the horizontal coherences in the natural local basis is indeed negative and equal to 
 \bea
-\Delta^{\infty} {\cal C}_{\rm h}&:=&-[{\cal C}_{\rm h}(\infty)-{\cal C}_{\rm h}(0)] \nn\\
&\!\underset{\hbar\omega|\beta_0|\gg1}{=}\!\!& -\!\sum_{m=-ns}^{ns} \frac{e^{-\hbar\omega m\beta_B}}{Z_{ns}(\beta_B)} \ln I_m <0,
\eea
 where $Z_{ns}(\beta_B):= \sum_{m=-ns}^{ns}e^{-m\hbar \omega \beta_B}$, $\beta_B$ is the bath inverse temperature, and 
 $I_m$ 
  is a growing function of $n$ and $s$ corresponding to the degeneracy of the $m^{\rm th}$ excited energy level.

One could suspect that, on the other hand, such negative contribution to the entropy production would be compensated by an increase of the variation of $-{\cal D}_{\rm th}$, $-\Delta^{\infty} {\cal D}_{\rm th} :=-[{\cal D}_{\rm th}(\infty)-{\cal D}_{\rm th}(0)]$. However, this is not the case. Indeed, $-\Delta^{\infty} {\cal D}_{\rm th}$ is also reduced compared to independent dissipation (see Appendix \ref{appss}). 

  \begin{figure}
\centering
\includegraphics[width=7cm, height=4.5cm]{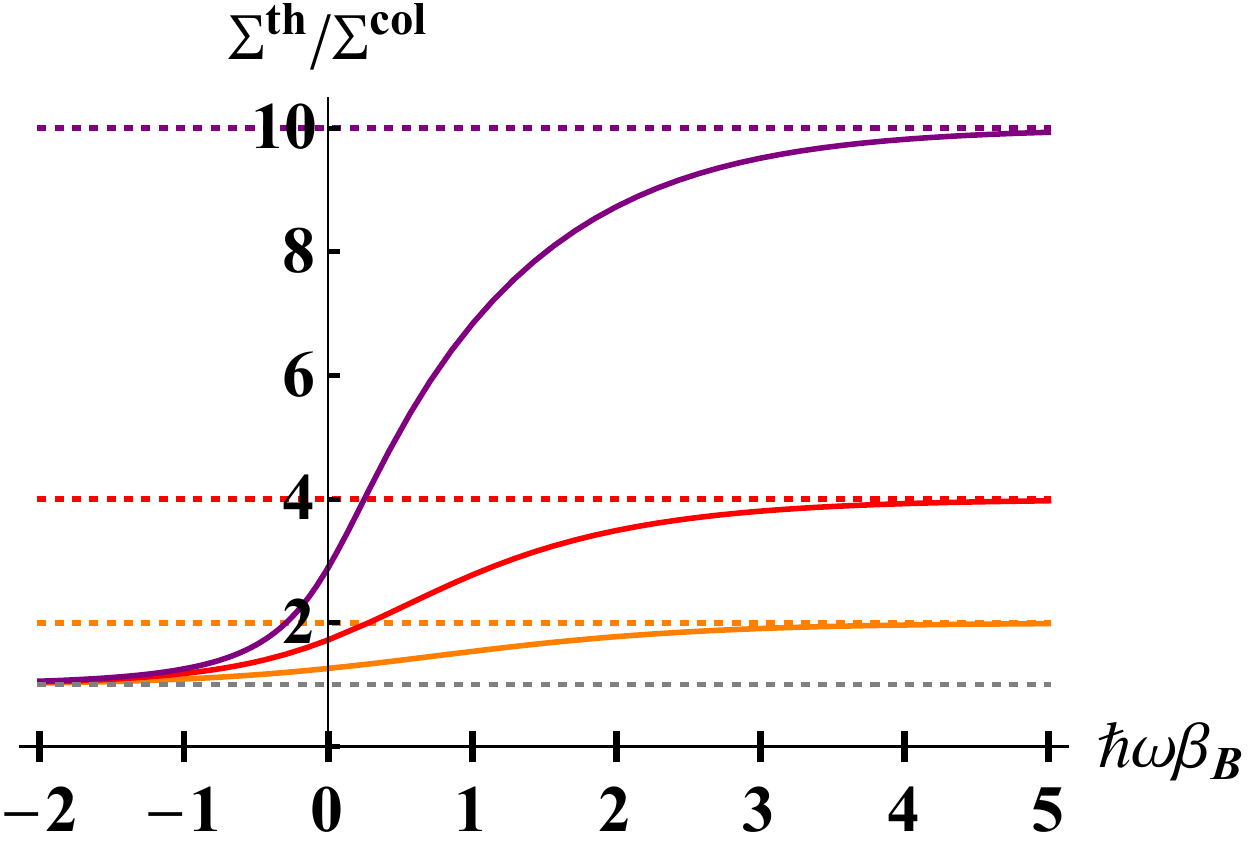}
\caption{Ratio of the entropy production with and without generation of horizontal coherences. The dissipation process with no generation of horizontal coherences corresponds to independent dissipation of each spin. The ensemble reaches the thermal equilibrium state $\rho^{\rm th}(\beta_B)$. The associated entropy production is denoted by $\Pi^{\rm th}$. The dissipation process with generation of horizontal coherences corresponds to collective dissipation of the spins. The ensemble reaches the equilibrium state $\rho^{\infty}_{\beta_0}(\beta_B)$ (see Appendix \ref{appss} for the detailed expression). The associated entropy production is denoted by $\Pi^{\rm col}$. The graph represents the plot of the ratio $\Pi^{\rm th}/\Pi^{\rm col}$ as a function of $\hbar\omega\beta_B$ for $\hbar\omega|\beta_0|\gg1$ and for ensembles containing $n=2$ (orange curve), $n=4$ (red curve), $n=10$ (purple curve) spins $s=1/2$. The associated dotted line emphasises the asymptotic value of each curve, equal to $n$, the number of spins in the ensemble. The grey dotted line indicates 1. The curves have been plotted using expression derived in \cite{bathinducedcoh}.}
\label{entrprod}
\end{figure}

  Therefore, the 
  negative contribution $-\Delta^{\infty} {\cal C}_{\rm h}$ promote a {\it reduction} of the entropy production and consequently of the irreversibility of the process, reenforced by the decrease of $-\Delta^{\infty} {\cal D}_{\rm th}$. This elucidates 
  the origin of the reduction of entropy production pointed out in \cite{bathinducedcoh}. As an illustration of the importance of the reduction, Fig. \ref{entrprod} displays for $\hbar\omega|\beta_0|\gg1$ the plots of the ratio $\Pi^{\rm th}/\Pi^{\rm col}$, where $\Pi^{\rm th}$ ($\Pi^{\rm col}$) is the entropy production {\it without} ({\it with}) generation of horizontal coherences, which corresponds to independent (collective) dissipation \cite{bathinducedcoh}. 
  One can see that for $\hbar\omega\beta_B >1$, the entropy production of the collective dissipation $\Pi^{\rm col}$ tends to be $n$ times smaller than $\Pi^{\rm th}$ (coinciding with the analytical results derived in \cite{bathinducedcoh}).
    %
%
As mentioned in the introduction, 
 irreversibility is known for degrading the performances of thermodynamic operations 
 \cite{Barato_2015, Gingrich_2016, Pietzonka_2016, Pietzonka_2018, Guarnieri_2019, Timpanaro_2019, Su_2019, Holubec_2018}. Therefore, the reduction of entropy production and irreversibility presented here might become useful to avoid such degradation.



\subsection{Negative velocity of the population convergence} 
\begin{figure}
\centering
\includegraphics[width=7cm, height=4cm]{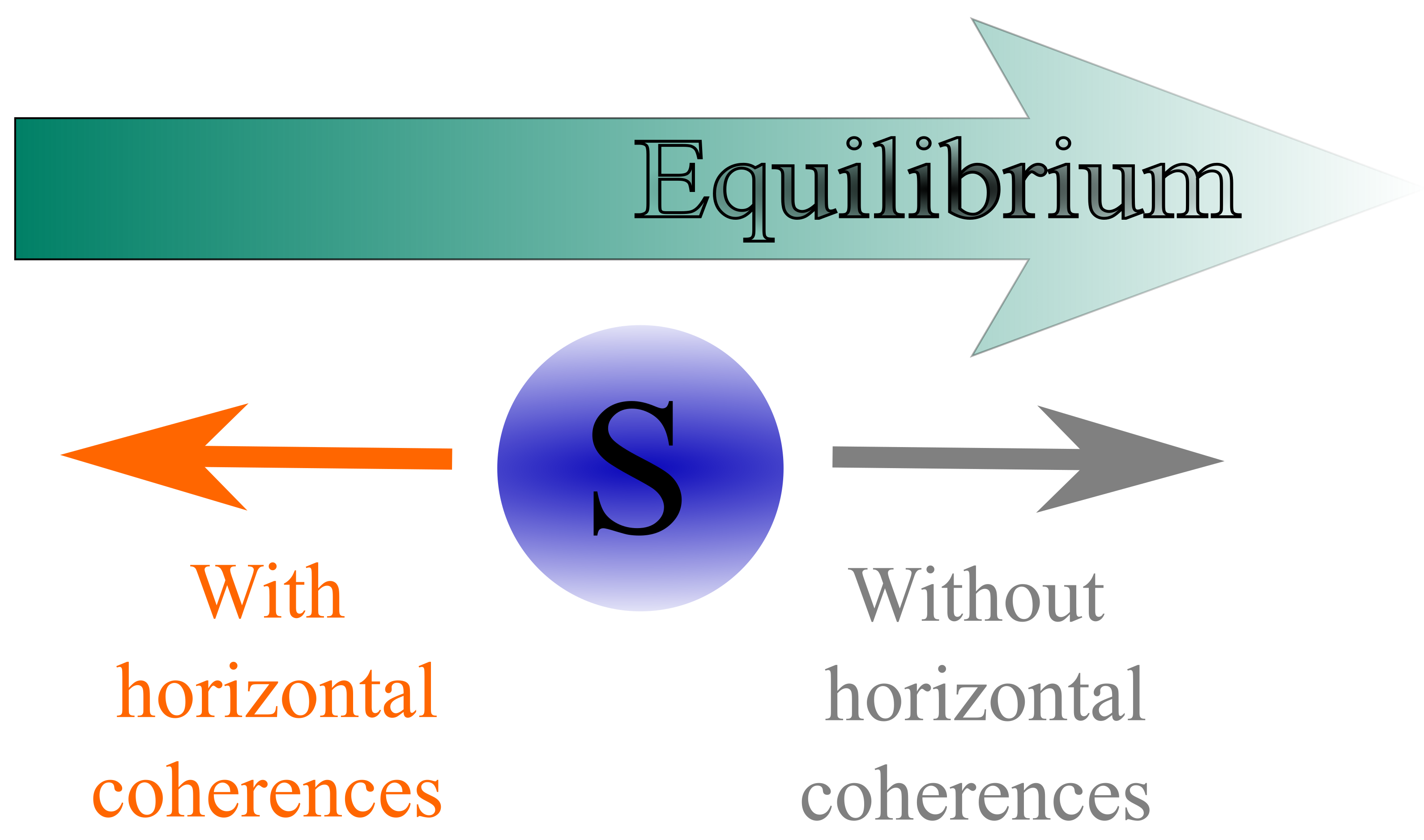}
\caption{{\bf Illustration of reversal of population convergence.} Without horizontal coherences, the populations naturally converge monotonically to the thermal equilibrium distribution. The presence of horizontal coherences can have an astonishing effect: reversal of the population convergence, resulting in populations going away from the thermal equilibrium distribution. Moreover, the ``fuel'' enabling this reversal of the natural population convergence is the horizontal coherences themselves. }
\label{divergence}
\end{figure}

As mentioned above, 
 $-\dot {\cal D}_{\rm th}$ corresponds to the velocity of convergence of the population distribution towards the thermal equilibrium distribution, and is also related to (minus) the time derivative of the diagonal free energy ${\cal F}_{\rm D}$. For non-degenerate systems, this velocity is always positive, corresponding the the expected monotonic population convergence to the thermal equilibrium distribution (or monotonic decrease of the diagonal free energy).
 By contrast, for degenerate systems, the coupled dynamics of the populations and horizontal coherences implies $\rho_{t+dt|_{\rm D}} \ne e^{dt {\cal L}}\rho_{t|_{\rm D}}$, which breaks down the guarantee of positivity of $-\dot {\cal D}_{\rm th}$ (as in \eqref{cnen} for $-\dot {\cal C}_{\rm h}$).
 Strikingly, this means that the population distribution can go away from the thermal equilibrium distribution, as illustrated in Fig. \ref{divergence}, or equivalently that the diagonal free energy can increase. Furthermore, we will see that this phenomenon can indeed be related to the heat flow reversal pointed out in \cite{heatflowreversal}, illustrated in Fig. \ref{reversal}. 

In the following we present a situation exhibiting such properties. 
We consider a system $S$ following the dynamics \eqref{me} and initially in a state of the form
\be\label{initialstate}
\rho_0=\rho^{\rm th}(\beta_0)+\chi,
\ee
where $\rho^{\rm th}(\beta_0)$ is the thermal state at inverse temperature $\beta_0$ and $\chi$ is an arbitrary hermitian matrix containing only off-diagonal terms (vertical and horizontal coherences) in the chosen basis $\{|n,i\ket\}$, so that $\rho_{0|_{\rm D}} = \rho^{\rm th}(\beta_0)$. Then, the velocity of the population convergence at initial times (small with respect to $\Gamma^{-1}$) is,
\bea\label{relationvelheat}
-\dot {\cal D}_{\rm th}
&=&-{\rm Tr}\dot \rho_{t=0} \big(\ln \rho^{\rm th}(\beta_0) - \ln \rho^{\rm th}(\beta_B)\big)\nn\\
&=& (\beta_0-\beta_B) \dot E_S.
\eea
The heat flow $\dot E_S$ can be written in the following form \cite{paperapptemp,autonomousmachines} (see Appendix  \ref{apphflow}),
\bea\label{mainhf}
\dot E_S  &=&\sum_{\omega>0}\omega G(\omega)\bra\ca\cad\ket_{\rho_t}\big(e^{-\omega\beta_B}-e^{-\omega/{\cal T}(\omega)}\big),\nn\\
 \eea
where ${\cal T}(\omega):= \omega \left(\ln\frac{\bra \ca\cad\ket_{\rho_t}}{\bra\cad\ca\ket_{\rho_t}}\right)^{-1}$ is the apparent temperature associated with the energy exchange $\omega$ \cite{paperapptemp,autonomousmachines}. In particular, for initial states of the form \eqref{initialstate} the inverse apparent temperatures can be rewritten as 
\be\label{cohinapptemp}
\frac{\omega}{{\cal T}(\omega)} = \omega\beta_0 +  \ln \frac{1+ c^{+}}{1+c^{-}}
\ee
where $c^{-}:= \bra \ca^{\dag} \ca \ket_{\chi}/\bra \ca^{\dag} \ca \ket_{\rho^{\rm th}(\beta_0)}$ and $c^{+}:=\bra \ca \ca^{\dag} \ket_{\chi}/\bra \ca \ca^{\dag} \ket_{\rho^{\rm th}(\beta_0)}$ constitute the contribution from the {\it horizontal coherences}, highlighted in \cite{paperapptemp,heatflowreversal,autonomousmachines}. When $\chi$ do not contain {\it horizontal coherences} we have $\bra\ca\cad\ket_{\chi}=\bra\cad\ca\ket_{\chi}=0$, implying ${\cal T}(\omega)=1/\beta_0$. Consequently, in absence of horizontal coherences, the population convergence velocity becomes
\bea
-\dot {\cal D}_{\rm th} &=& (\beta_0-\beta_B)\sum_{\omega>0}\omega G(\omega)\bra\ca\cad\ket_{\rho_t}\nn\\
&&\hspace{2cm}\times\big(e^{-\omega\beta_B}-e^{-\omega\beta_0}\big),
\eea
which is always positive for any value of $\beta_0$ and $\beta_B$. 
However, in presence of horizontal coherences the apparent temperatures ${\cal T}(\omega)$ can be risen beyond or lowered below $1/\beta_B$ \cite{heatflowreversal}, inverting the role of hottest and coldest system and resulting in changing the {\it sign} of the heat flow \eqref{mainhf}. Consequently, from \eqref{relationvelheat}, the population convergence velocity $-\dot {\cal D}_{\rm th}$ becomes {\it negative}. 
 It is shown in \cite{heatflowreversal} that such heat flow reversals are always achievable for $\beta_0$ not too far from $\beta_B$. Note that the total heat exchanged between the initial and final time (when reaching the equilibrium state) can also be inverted, leading to $(\beta_0-\beta_B)\Delta E_S\leq 0$. 
 
We just showed that the velocity of the population convergence $-\dot {\cal D}_{\rm th}$ can become negative thanks to horizontal coherences, and that heat flow reversals is one of its observable consequences. 
In the next paragraph we go further: we show formally that heat flow reversals are {\it powered} by horizontal coherences (illustrated in Fig. \ref{reversal}).  \\

\subsection{Complementarity of horizontal coherences and heat flow reversal} \label{secresource}
\begin{figure}
\centering
\includegraphics[width=7cm, height=7.5cm]{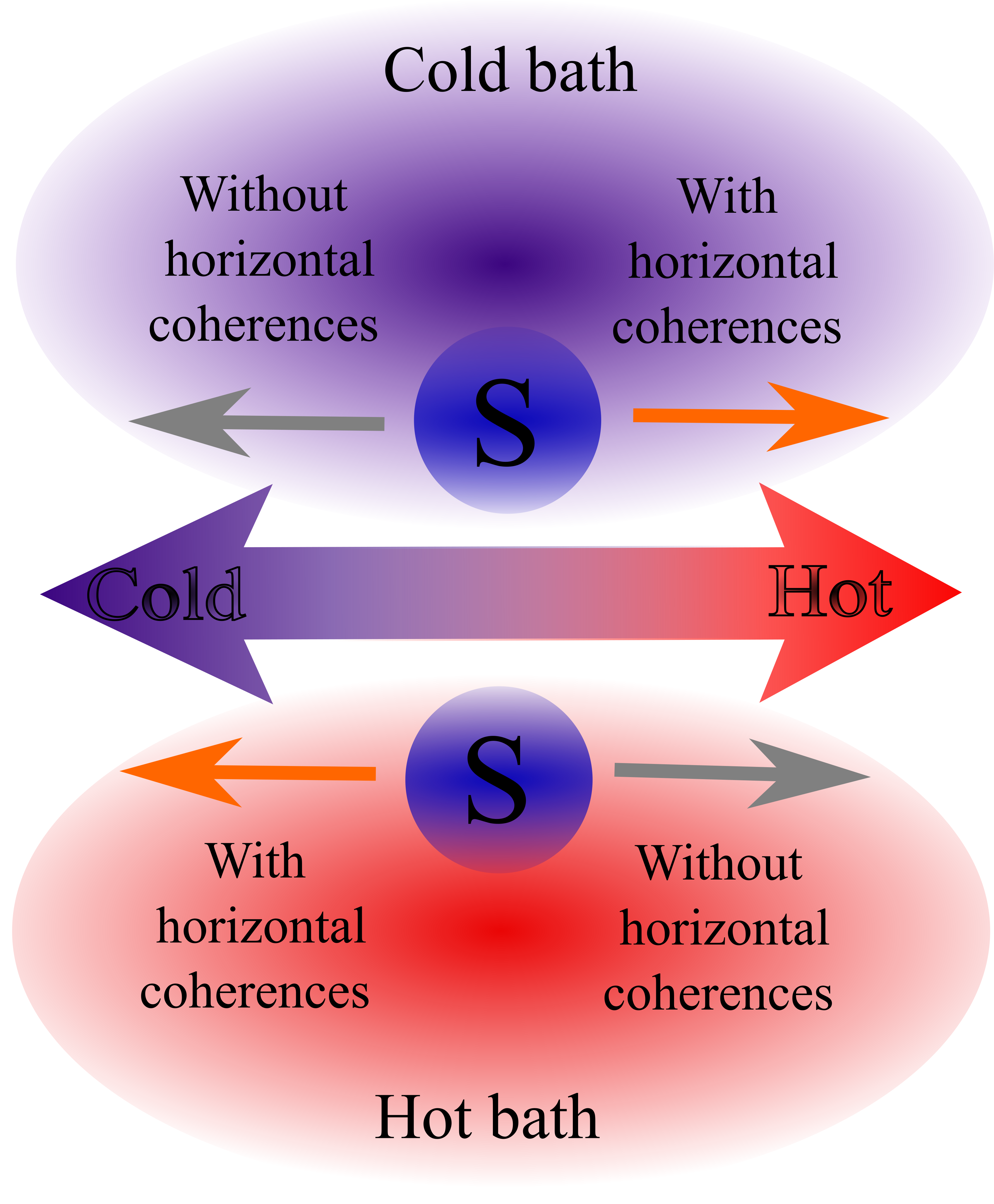}
\caption{{\bf Illustration of heat flow reversal.} The system $S$ can gain energy while interacting with a colder bath thanks to horizontal coherences acting as a ``fuel'' for heat flow reversal. Conversely, the system $S$ can lose energy while interacting with a hotter bath, again thanks to horizontal coherences.}
\label{reversal}
\end{figure}

We showed that both $-\dot {\cal C}_{\rm h}$ and $-\dot {\cal D}_{\rm th}$ can become negative.
  However, there is a restriction: the sum of $-\dot {\cal C}_{\rm h}$ and $-\dot {\cal D}_{\rm th}$ has to be always positive. Indeed, a variation of $- {\cal C}_{\rm h}(t) - {\cal D}_{\rm th}(t)$ between instants of time $t$ and $t'>t$  gives
\bea\label{globalpos}
&-& \Delta {\cal C}_{\rm h} - \Delta {\cal D}_{\rm th} \nn\\
&&= - \left[S(\rho_{t'|_{\rm BD}}|\rho^{\rm th}(\beta_B)) - S(\rho_{t|_{\rm BD}}|\rho^{\rm th}(\beta_B))\right]\nn\\
&&= - \Big[S(e^{(t'-t){\cal L}}\rho_{t|_{\rm BD}}|e^{(t'-t){\cal L}}\rho^{\rm th}(\beta_B)) \nn\\
&&\hspace{3.6cm}- S(\rho_{t|_{\rm BD}}|\rho^{\rm th}(\beta_B))\Big]\nn\\
&&\geq 0,
\eea
where the identity \eqref{equalitybis} and the contractivity of the relative entropy under completely positive and trace preserving maps \cite{Lindblad_1975,Petruccione_Book} were used in third and forth lines, respectively. The above inequality implies in particular that the time derivative of the sum is also always positive,
\bea
&-& \dot {\cal C}_{\rm h} - \Dot {\cal D}_{\rm th} \geq 0.
\eea
The physical meaning of the inequality \eqref{globalpos} appears after re-writing the variation of ${\cal D}_{\rm th}(t)$ between the initial time $t=0$ and any arbitrary later time $t$ as 
\be\label{denergy}
 - \Delta {\cal D}_{\rm th} = (\beta_0-\beta_B)\Delta E_S - S(\rho_{t|_{\rm D}}|\rho_{0|_{\rm D}}),
\ee
where $\Delta E_S= {\rm Tr}H_S(\rho_t-\rho_0)$ is the associated variation of energy of $S$. Note that \eqref{denergy} is valid for populations initially thermally distributed, $\rho_{0|_{\rm D}}=\rho^{\rm th}(\beta_0)$. Injecting \eqref{denergy} in the inequality \eqref{globalpos} one obtains
\be\label{ineqresource}
- \Delta {\cal C}_{\rm h} +(\beta_0-\beta_B)\Delta E_S - S(\rho_{t|_{\rm D}}|\rho_{0|_{\rm D}})\geq 0.
\ee
The quantity $(\beta_0-\beta_B)\Delta E_S$ is always positive for initial states without horizontal coherences. However, as shown above 
and in \cite{heatflowreversal}, the presence of initial horizontal coherences can reverse the heat flow $\dot E_S$ and the finite heat exchange $\Delta E_S$, implying $(\beta_0-\beta_B)\dot E_S<0$ and $(\beta_0-\beta_B)\Delta E_S<0$. From \eqref{ineqresource}, a reversal of finite heat exchange implies
\bea\label{nencohcost}
- \Delta {\cal C}_{\rm h} 
&\geq& -(\beta_0-\beta_B)\Delta E_S>0,
\eea
which means a strict {\it consumption} of horizontal coherences. Conversely, when horizontal coherences are not consumed, meaning that $- \Delta {\cal C}_{\rm h}=0$ (or even $- \Delta {\cal C}_{\rm h}< 0$), one has necessarily
\be\label{energycost}
(\beta_0-\beta_B)\Delta E_S \geq S(\rho_{t|_{\rm D}}|\rho_{0|_{\rm D}})\geq 0,
\ee
so that no reversal of heat exchange can happen. This shows explicitly that reversal of heat exchange is powered by horizontal coherences (illustration in Fig. \ref{reversal}). 
Additionally, the inequalities \eqref{globalpos} and \eqref{ineqresource} can be seen as kind of Landauer principles for horizontal coherences: they impose a lower bound on the energetic cost of creation of horizontal coherences. Namely, the creation of horizontal coherences is conditioned by
\be\label{energycost2}
(\beta_0-\beta_B)\Delta E_S\geq \Delta {\cal C}_{\rm h}> 0,
\ee
 implying an energetic cost paid in the form of heat (or ``natural'' heat exchange). 

Finally, it is also interesting to look at the time derivative version of \eqref{ineqresource}, giving at initial times ($0\leq t\ll \Gamma^{-1}$),
\be\label{tderivativeversion}
- \dot {\cal C}_{\rm h} +(\beta_0-\beta_B)\dot E_S \geq 0.
\ee
One can obtain straightforwardly the time-derivative analogue of the above inequalities \eqref{nencohcost}, \eqref{energycost}, and \eqref{energycost2}.
The absence of the term $ S(\rho_{t|_{\rm D}}|\rho_{0|_{\rm D}})$ (equal to zero for $0\leq t\ll \Gamma^{-1}$) in \eqref{tderivativeversion} provides the following insight: initially, both heat flow reversal and creation of horizontal coherences do not have extra costs in horizontal coherences and energy, respectively. However, as time passes, an extra cost corresponding to $S(\rho_{t|_{\rm D}}|\rho_{0|_{\rm D}})\geq 0$ is required,  as shown by \eqref{nencohcost} and \eqref{energycost2}. 
In particular, this explains why reversals of finite heat exchanges require larger amount of initial coherences than heat flow reversals, as pointed out and discussed in \cite{heatflowreversal}.\\

\section{Thermal and a-thermal operations} 
We now show that the above results are indeed valid in a broader context. The following considerations are inspired from \cite{Santos_2019}, with new results related to horizontal coherences.

 Let's consider the unitary interaction of our system of interest $S$ with an other system $B$ between two instant of times $t_i$ and $t_f$. Importantly, in the remainder of the paper $B$ is not restricted to baths but can be of any size, even an elementary single quantum system. 
  We denote by $U$ the associated unitary transformation (acting on both $S$ and $B$), and by $\rho_{X, t}$ the density operator of the system $X$ (standing for $S$, $B$, or $SB$) at an arbitrary instant of time $t$. Note that in principle there is also no restriction on the strength of the coupling between $S$ and $B$ (not anymore limited to weak coupling as in \eqref{me}). 
 
 We call {\it a-thermal}, operations satisfying the conditions of initial separability, energy conservation, and stationarity of $\rho_{B,t_i}$, expressed respectively by $\rho_{SB, t_i}=\rho_{S, t_i}\rho_{B,t_i}$, $[U,H_S+H_B]=0$, and $[H_B,\rho_{B,t_i}]=0$. If one asks additionally the initial state $\rho_{B,t_i}$ to be thermal, the operation belongs to the well-known set of {\it thermal operations} \cite{Janzing_2000,Horodecki_2013,Meer_2017,Muller_2018,Perry_2018}.
 Interestingly, the only conditions defining a-thermal operations guarantee the validity of most of the above results. More precisely, under the energy conservation condition, one can show (see Appendix\ref{appbroad}) that the evolution of the vertical coherences is closed (in the sense mentioned above). Thus, one can simply repeat the argument \eqref{mainposen} (slightly adapted to finite evolution, see Appendix \ref{appbroad})
 and show that the vertical coherences always decrease,
\be
-\Delta {\cal C}_{\rm v} = -[{\cal C}_{\rm v}(t_f)-{\cal C}_{\rm v}(t_i)] \geq 0.
\ee
The are two key points. First, $\Lambda$, denoting the reduced evolution of $S$, is a completely positive trace-preserving map (thanks to the initial separability and the unitarity of the global evolution \cite{Petruccione_Book}), which guarantees the contraction of the relative entropy. Second, the closed evolution of the vertical coherences, which guarantees 
\be\label{mainbdid}
\rho_{S,t_f|_{\rm BD}}=\Lambda \rho_{S,t_i|_{\rm BD}}.
\ee
By contrast, the mixing of the horizontal coherences' dynamics with the populations' dynamics implies 
\be
\rho_{S,t_f|_{\rm D}}\ne\Lambda \rho_{S,t_i|_{\rm D}},
\ee
which, repeating the argument in \eqref{cnen}, breaks down the guarantee of positivity of $-\Delta {\cal C}_{\rm h}$. In Appendix \ref{appbroad}, some explicit conditions (relying on degenerate transitions, Fig. \ref{energylevels})  are pinpointed in order to have effectively $-\Delta {\cal C}_{\rm h}<0$.

Assuming the existence of a thermal equilibrium state $\rho_{S}^{\rm th}(\beta_B)$, which is guaranteed for instance when $B$ is initially in a thermal state, 
one can follow the definitions introduced in the beginning of the paper for the entropy production $\Pi$ and the measure of population distance to the equilibrium distribution ${\cal D}_{\rm th}$. 
 Then, for the same reason of non-closure of the populations' dynamics, the guarantee of  positivity of $-\Delta {\cal D}_{\rm th}$ is broken. Again, in Appendix \ref{appclaw}, some conditions are mentioned for having divergence of the populations from the equilibrium distribution, $-\Delta {\cal D}_{\rm th}<0$. 

Furthermore, as above, 
one can show with the same arguments used for vertical coherences that  
\be\label{generalisation1}
-\Delta {\cal C}_{\rm h} -\Delta {\cal D}_{\rm th} \geq 0.
\ee 
As in the previous Section, this can be thought as a kind of Landauer principle for the horizontal coherences. Even though in general $-\Delta {\cal D}_{\rm th}$ cannot be simply related to energy exchanges as in \eqref{denergy}, it is still associated to the variation of diagonal free energy,  $-\Delta {\cal D}_{\rm th}=-\beta_B\Delta {\cal F}_{\rm D}$, and still represents the convergence of the populations towards the thermal equilibrium distribution. 
Assuming for instance a situation where $U$ generates horizontal coherences, one obtains 
\be\label{generalisation2}
 -\Delta {\cal D}_{\rm th} \geq \Delta {\cal C}_{\rm h} >0,
 \ee
  which represents the necessary cost in ``population gradient'' for creating horizontal coherences. In particular, $\rho_{S,t_i|_{\rm D}}$ has to be far enough from $\rho_{S}^{\rm th}(\beta_B)$. 
Conversely, having $-\Delta {\cal D}_{\rm th} <0$, meaning a {\it reversal} of the natural tendency of convergence of the population to the thermal equilibrium distribution, necessarily requires consumption of horizontal coherences, 
\be\label{generalisation3}
 -\Delta {\cal C}_{\rm h} \geq \Delta {\cal D}_{\rm th} >0,
 \ee
 and therefore initial states containing horizontal coherences. 
This result is a generalisation of \eqref{nencohcost} and of the heat flow reversal \cite{heatflowreversal}.

Interestingly, the above inequalities can be extended to equalities. Based on the observation that the operation of global block diagonalisation commute with $U$, $U\rho_{SB, t|_{\rm BD}}U^{\dag} = \left(U\rho_{SB,t}U^{\dag}\right)_{|_{\rm BD}}$ (see Appendix \ref{appclaw}), where $\rho_{SB,t}$ stands for the density matrix of the ensemble $SB$, 
one can show 
a {\it conservation law} of vertical coherences (already obtained in \cite{Santos_2019} in a different form),
\be\label{conservationen1}
\Delta {\cal C}_{\rm v}^{SB}=0.
\ee
In the above conservation law we defined ${\cal C}_{\rm v}^{SB}(t):=S(\rho_{SB,t|_{\rm BD}}) - S(\rho_{SB,t})$ as the extension of the relative entropy of vertical coherences to the ensemble $SB$.
The conservation law can be alternatively written as
\be\label{conservationen2}
-\Delta {\cal C}^S_{\rm v}-\Delta {\cal C}^B_{\rm v} = {\cal C}_{\rm c, v}^{SB}(t_f).
\ee
The introduced superscripts $S$, $B$, or $SB$ refer to the corresponding systems and $ {\cal C}_{\rm c, v}^{SB}(t)$ is the {\it correlated} vertical coherences introduced in \cite{Tan_2016} as 
\be\label{defccoh}
{\cal C}_{\rm c, v}^{SB}(t):= {\cal C}_{\rm v}^{SB}(t)- {\cal C}_{\rm v}^{S}(t)- {\cal C}_{\rm v}^{B}(t)\geq0.
\ee
${\cal C}_{\rm c, v}^{SB}(t)$ represents the vertical coherences present in $SB$ fruits of correlations between $S$ and $B$. Therefore, the conservation law \eqref{conservationen2} implies in particular that the vertical coherences consumed in $S$ is either transferred to $B$ or to correlated vertical coherences ${\cal C}_{\rm c, v}^{SB}(t_f)$ between $S$ and $B$.
Indeed, this conservation law is surprising from the perspective of decoherence theory in open quantum systems \cite{Petruccione_Book} which taught us that coherences are destroyed by the bath. This is because open quantum systems theory is mainly concerned about the reduced system's dynamics and therefore changes in the bath state are mostly ignored.

By contrast, horizontal coherences {\it are not conserved}. This is one more aspects of the fundamental difference between vertical and horizontal coherences. However, one can derive a conservation law for horizontal coherences when contributions from the population convergence is included. Namely, one can show (see Appendix \ref{appclaw}), 
\be\label{conservationnen1}
 \Delta {\cal C}_{\rm h}^{SB}+\Delta {\cal D}_{\rm th}^{SB}=0,
  \ee
  which establishes that, globally, the reversal of population convergence is exactly equal to the consumption of horizontal coherences. In the above identity we introduced ${\cal C}_{\rm h}^{SB}(t):=S(\rho_{SB,t|_{\rm D}}) - S(\rho_{SB,t|_{\rm BD}})$ and $-{\cal D}_{\rm th}^{SB}(t):= -S[\rho_{SB,t}|\rho_{SB}^{\rm th}(\beta_B)]$, where $\rho_{SB}^{\rm th}(\beta_B)$ is the global thermal state at inverse temperature $\beta_B$ and is a steady state.  
  Conversely, generation of horizontal coherences (still globally) is exactly compensated by population convergence (which inherently carries an energetic cost). Additionally, the conservation law \eqref{conservationnen1} can be used to show that the generation of horizontal coherences affects the energy exchanges (see Appendix \ref{appclaw}), recovering in a straightforward way observations of \cite{bathinducedcohTLS,bathinducedcoh}. By contrast, when no horizontal coherences are consumed or generated, one has necessarily $\Delta {\cal D}_{\rm th}^{SB}=0$, implying that globally, the dynamics of the populations is restricted to a region of equidistant points to the equilibrium distribution.
  
  The conservation law \eqref{conservationnen1} can alternatively be expanded as
\be\label{conservationnen2}
-\Delta {\cal C}^S_{\rm h} -\Delta {\cal D}^S_{\rm th} -\Delta {\cal C}_{\rm h}^B-\Delta {\cal D}^B_{\rm th} = {\cal C}_{\rm c, h}^{SB}(t_f) + {\cal D}_{\rm c, th}^{SB}(t_f),
\ee
where the superscripts $S$, $B$, and $SB$ refer to the corresponding systems. The quantities ${\cal C}_{\rm c, h}^{SB}(t_f)$ and ${\cal D}_{\rm c, th}^{SB}(t_f)$ are respectively the horizontal global coherences and the distance of the global populations to the thermal equilibrium distribution $\rho_{SB}^{\rm th}(\beta_B)$ stemming from correlations between $S$ and $B$. Both ${\cal C}_{\rm c, h}^{SB}(t_f)$ and ${\cal D}_{\rm c, th}^{SB}(t_f)$ are defined in a similar way as in \eqref{defccoh} and are positive. The conservation laws \eqref{conservationnen1} and \eqref{conservationnen2} can be seen as extensions of \eqref{generalisation1}, \eqref{generalisation2}, and \eqref{generalisation3} in form of equalities. Eq. \eqref{conservationnen2} means that the horizontal coherences consumed in $S$ and $B$ plus the steps towards $\rho_{S}^{\rm th}(\beta_B)={\rm Tr}_B \rho_{SB}^{\rm th}(\beta_B)$ and $\rho_{B}^{\rm th}(\beta_B)={\rm Tr}_S \rho_{SB}^{\rm th}(\beta_B)$ is recovered in the final correlated horizontal coherences and distance to $\rho_{SB}^{\rm th}(\beta_B)$.

 Note that the time derivative version of the results of this section are valid only upon a divisibility property of $U$. Namely, for any $t\in[t_i;t_f]$, $U$ is divisible in two unitary evolutions $U_{t_i,t}$ and $U_{t,t_f}$ which are both energy conservatives.

\section{Concluding remarks}


The entropy production of a degenerate system can be split in three contributions. For bath-driven dissipation and thermal (or a-thermal) operations, the first contributions is always positive and is related to the consumption of vertical coherences. The second contribution stems from horizontal coherences. Differently from the first one, it can be either positive or negative, which is shown to be associated respectively to consumption or generation of horizontal coherences. In particular, this explains the origin of the entropy reduction pointed out in \cite{bathinducedcoh}.
Finally, the third contribution stems from the convergence of the populations to the thermal equilibrium distribution, and is related to the variation of the diagonal free energy. While this contribution is always positive in the absence of horizontal coherences, meaning that the populations always tends to the thermal equilibrium distribution, this natural tendency can be reversed thanks to horizontal coherences. This phenomenon is associated to an other negative contribution to the entropy production and is the origin of the heat flow reversal introduced in \cite{heatflowreversal}. The cost for this inversion is paid in horizontal coherences. Conversely, the generation of horizontal coherences is paid in heat or ``population gradient''. 
Finally, a global conservation law is recovered for vertical coherences, whereas 
 horizontal coherences are shown not to be conserved. Nevertheless, the sum of the horizontal coherences plus the population convergence rate is globally conserved. 



The phenomena pointed out in this paper rely on the degeneracy of $S$ and the existence of degenerate transitions (Fig. \ref{energylevels}). In such conditions the steady state of $S$ is not unique and therefore can be thought as not ``fully thermalised'', which can help to understand intuitively some of the above phenomena.
Regarding experimental implementations, degenerate transitions were realised in multi-level atoms for instance in \cite{Xia_1996}, and also appear in diverse contexts (superradiance and bath-induced entanglement) in ensembles of subsystems collectively coupled to a common bath, with some experimental realisations for instance in \cite{Krauter_2011, Norcia_2018,Devoe_1996, Raimond_1982}. 
Moreover, the exact degeneracy of $S$ is indeed not necessary. In Appendix \ref{neardegenerate}, it is shown that energy gaps  of order $\delta$ in the system's spectrum are initially not resolved by the bath when $\delta$ is smaller than the dissipation decay rate. Then, at least for times much smaller than $\delta^{-1}$, which can be still enough for significant dissipation of $S$ and even to reach a non-thermal steady state, the system appears as degenerate from the point of view of the bath and the phenomena described throughout the paper can take place. 
In particular, such situations include ensembles of non-interacting or even weakly interacting subsystems with inhomogeneities smaller than the dissipative rate, which can be achieved for instance with atomic cloud in ``bad'' optical cavity \cite{Barberena_2019,Norcia_2016,Norcia_2018}. 
 
Furthermore, motivating more research in this direction, our results seem to indicate that, in the context of Markovian bath dissipations and thermal operations, thermodynamics departs from classical behaviour thanks to horizontal coherences and the associated negative contributions. 
An interesting parallel can be made with a recent study \cite{Smirne} where the authors establishes that a statistics obtained from measurements on a quantum system is genuinely quantum (meaning that it does not admit classical description) if and only if the underlying dynamics generates coherences which can be subsequently turned into populations -- precisely the mechanism associated to the negative contributions to entropy production. 
In addition, on top of the effects on the ongoing thermodynamics already reported throughout the paper, one can show that negative contributions to entropy production can modify the performances of thermal machines (see Appendix \ref{appeffecttm}).
It is also interesting to note that, since thermal operations can indeed {\it generate} horizontal coherences out of incoherent states -- contrarily to what happens with vertical coherences \cite{Lostaglio_2015, Streltsov_2017} -- the resource theory of vertical and horizontal coherences do not share the same set of free operations. This emphasises again the special statue of horizontal coherences and suggests the possibility of introducing a specific resource theory for horizontal coherences. 

Beyond that, it would also be interesting to investigate what could be the role of horizontal coherences in work fluctuation reductions achieved through collective operations \cite{Llobet_2019}, violation of work fluctuation relations in processes generating coherences \cite{Miller_2019}, thermodynamic uncertainty relations \cite{Barato_2015,Gingrich_2016,Guarnieri_2019,Timpanaro_2019}, and dissipated work in non-adiabatic driving \cite{Francica_2019}.

\acknowledgements
C.L.L. would like to thank Gabriel T. Landi for interesting discussions. 
This  work  is  based  upon  research  supported  by  the
South  African  Research  Chair  Initiative  of  the  Department  of  Science  and  Technology of the Republic of South Africa, the National  Research Foundation of the Republic of South Africa, and National Institute for Theoretical Physics (NITheP) of the Republic of South Africa.

\appendix
\numberwithin{equation}{section}

 \section{Dissipation of near degenerate systems}\label{neardegenerate}
 In the main text we assume a perfectly degenerate system $S$. In this paragraph we show that this requirement can be significantly relaxed. We consider that the Hamiltonian of $S$ is non-degenerate and of the form 
 \be\label{neardegen}
 H_S= \sum_n \sum_{i=1}^{l_n} e_{n,i}|n,i\ket\bra n,i|.
 \ee
  However, we assume that for all $n$ and all $i,i'$, the energy gap $|e_{n,i}-e_{n,i'}|$ is at most of the order of magnitude $\delta$, whereas for all $n$, $m$, $i$, and $j$, $|e_{n,i}-e_{m,j}|$ is of the order of magnitude $\omega$, with $\omega\gg\delta$. In the following we refer to these properties as near degeneracy. Note that several systems, from multi-level atoms containing some close energy levels to ensembles of interacting subsystems with inhomogeneities, are near degenerate. Indeed, in the later situation, one can always decompose the Hamiltonian of the ensemble as $H_S= H_0 + H_{\rm inh}+H_{\rm int}$, where $H_0$ denotes the sum of the local free Hamiltonians of each subsystem, $H_{\rm inh}$ corresponds to inhomogeneities representing potential small differences between the subsystems (like different energy transitions), and $H_{\rm int}$ stands for interaction between the subsystems. Using for instance the degenerate perturbation theory \cite{LandauLifshitz_Book}, one can easily see that $H_S$ can indeed be re-written in the form \eqref{neardegen} satisfying the near degeneracy criteria. 
  In other words, near degeneracy represents a more realistic situation since it is expected that any perfectly degenerate system looses its perfect degeneracy due to unavoidable interactions and perturbations from the environment, as just shown.

  For such a near degenerate system $S$, the Markovian master equation is in principle different from \eqref{me} because the levels $|n,i\ket$, $1\leq i\leq l_n$ are not degenerate. Furthermore, the secular approximation \cite{Petruccione_Book, Cohen_Book} is not valid anymore if $\delta$ is of the order or smaller than $g^2\tau_c$, where $\tau_c$ denotes the bath coherence time and $g$ is the system-bath coupling strength. In the following, we show that indeed one can recover a master equation of the form \eqref{me} when $\delta\ll g^2\tau_c\ll \omega$. Starting from the Born and Markov approximations (valid for $g\tau_c\ll1$) \cite{Petruccione_Book, Cohen_Book}, the reduced dynamics of $S$ before applying the secular approximation is
  \bea\label{derme}
  \dot \rho_t &=& \int_0^{\infty} du {\rm Tr}_B \Big[V(t-u)\rho_t\rho_B V(t) - V(t)V(t-u)\rho_t\rho_B\Big] \nn\\
  &&+ {\rm h.c.}\nn\\
  &=& \sum_{m,n,m',n'}\sum_{i,i',j,j'} \Gamma(e_{m,j}-e_{n,i})\nn\\
  &&\times e^{-i(e_{m,j}-e_{n,i}-e_{m',j'}+e_{m,j})t}A_{n,i,m,j}A^{*}_{n',i',m',j'}\nn\\
  &&\times\Big(|n,i\ket\bra m,j|\rho_t|m',j'\ket\bra n',i'|\nn\\
  &&\hspace{1cm}-|m',j'\ket\bra n',i'|n,i\ket\bra m,j|\rho_t\Big) + {\rm h.c.}
  \eea
  where $V(t)$ is the coupling Hamiltonian in the interaction picture (with respect to $H_S$), 
 the coefficients $A_{n,i,m,j}:= \bra n,i|A_S|m,j\ket$ are the amplitudes of transitions, and 
 the sum $\sum_{i,i',j,j'}$ is a short notation for $\sum_{i=1}^{l_n}\sum_{i'=1}^{l_{n'}}\sum_{j=1}^{l_m}\sum_{j'=1}^{l_{m'}}$. There are three kinds of terms in the above sum \eqref{derme}. Terms such that $|e_{m,j}-e_{n,i}-(e_{m',j'}-e_{n',i'})|$ is of order $\omega$, terms such that $|e_{m,j}-e_{n,i}-(e_{m',j'}-e_{n',i'})|$ is of order $\delta$, and terms such that $e_{m,j}-e_{n,i}-(e_{m',j'}-e_{n',i'})=0$. For the first group of terms, assuming $\omega \gg g^2\tau_c$, we can apply the secular approximation: the phase $e^{-i(e_{m,j}-e_{n,i}-e_{m',j'}+e_{n',i'})t}$ evolves much faster than the evolution timescale of $\rho_t$ (of order $(g^2\tau_c)^{-1}$) so that the average contribution of such terms is zero. The second and third group of terms can be put together as follows. For simplicity, we detail the procedure for a pair of transitions $\{|m,j\ket, |m,j'\ket\} \rightarrow \{|n,i\ket, |n,i'\ket\}$, but this can be extended straightforwardly to all remaining transitions. There are four terms associated with this pair of transitions:
 \begin{widetext}
 \bea\label{4terms}
 &&e^{-i(e_{m,j}-e_{n,i}-e_{m,j'}+e_{n,i'})t}\Gamma(e_{m,j}-e_{n,i}) A_{n,i,m,j}A^{*}_{n,i',m,j'}\big(|n,i\ket\bra m,j|\rho_t|m,j'\ket\bra n,i'|-|m,j'\ket\bra n,i'|n,i\ket\bra m,j|\rho_t\big)+ {\rm h.c.}\nn\\
 &&+e^{-i(e_{m,j'}-e_{n,i'}-e_{m,j}+e_{n,i})t}\Gamma(e_{m,j'}-e_{n,i'}) A_{n,i',m,j'}A^{*}_{n,i,m,j}\big(|n,i'\ket\bra m,j'|\rho_t|m,j\ket\bra n,i|-|m,j\ket\bra n,i|n,i'\ket\bra m,j'|\rho_t\big)+ {\rm h.c.}\nn\\ 
 &&+\Gamma(e_{m,j}-e_{n,i}) |A_{n,i,m,j}|^2\big(|n,i\ket\bra m,j|\rho_t|m,j\ket\bra n,i|-|m,j\ket\bra n,i|n,i\ket\bra m,j|\rho_t\big)+ {\rm h.c.}\nn\\
 &&+\Gamma(e_{m,j'}-e_{n,i'}) |A_{n,i',m,j'}|^2\big(|n,i'\ket\bra m,j'|\rho_t|m,j'\ket\bra n,i'|-|m,j'\ket\bra n,i'|n,i'\ket\bra m,j'|\rho_t\big)+ {\rm h.c.}.
 \eea
 \end{widetext}
 Assuming $\delta\tau_c\ll1$ (which is automatically satisfied if $\delta\ll g^2\tau_c$), we have the following relation,
 \bea 
 \Gamma(\omega+\delta)-\Gamma(\omega) &=& \int_0^{\infty} du(e^{i(\omega+\delta)u}-e^{i\omega u}){\rm Tr}\rho_BA_B(u)A_B\nn\\
 &\simeq &\int_0^{\tau_c}  du(e^{i(\omega+\delta)u}-e^{i\omega u}){\rm Tr}\rho_BA_B(u)A_B\nn\\
 &=& \int_0^{\tau_c}  due^{i\omega u}[i\delta u +{\cal O}(\tau_c^2\delta^2)]{\rm Tr}\rho_BA_B(u)A_B\nn\\
 &=& \delta \frac{\partial \Gamma(\omega)}{\partial \omega} + {\cal O}[\tau_c^2\delta^2|\Gamma(\omega)|].
 \eea
 The derivative of $\Gamma(\omega)$ is of the order $g^2\tau_c^2$,
 \bea
 \Big|\frac{\partial \Gamma(\omega)}{\partial \omega}\Big| &\leq & \int_0^{\tau_c} u |{\rm Tr}\rho_BA_B(u)A_B| du \nn\\
 &&\sim \int_0^{\tau_c} u g^2 du=\frac{1}{2}g^2\tau_c^2.\nn\\
 \eea
 Thus, all together we obtain $|\Gamma(\omega+\delta)-\Gamma(\omega)|\sim g^2\tau_c^2 \delta \ll g^2\tau_c \sim |\Gamma(\omega)|$.
 Furthermore, considering times $t$ such that $\delta  t\ll 1$, one can approximate the above phases by 1, $e^{-i(e_{m,j}-e_{n,i}-e_{m,j'}+e_{n,i'})t}\simeq 1$. Then, the sum \eqref{4terms} of the 4 terms factorises in the following form
 \begin{widetext}
 \bea
 \eqref{4terms} &=& \Gamma(e_{m,j}-e_{n,i})\Big[\big(A_{n,i,m,j}|n,i\ket\bra m,j|+A_{n,i',m'j'}|n,i'\ket\bra m,j'|\big)\rho_t\big(A^{*}_{n,i,m,j}|m,j\ket\bra n,i|+A^{*}_{n,i',m,j'}|m,j'\ket\bra n,i'|\big)-\nn\\
 &&\hspace{0.5cm}-\big(A^{*}_{n,i,m,j}|m,j\ket\bra n,i|+A^{*}_{n,i',m,j'}|m,j'\ket\bra n,i'|\big) \big(A_{n,i,m,j}|n,i\ket\bra m,j|+A_{n,i',m'j'}|n,i'\ket\bra m,j'|\big)\rho_t\Big]+ {\rm h.c.}.
 \eea
 \end{widetext}
 Repeating this procedure for all the transitions $|m,j\ket\rightarrow|n,i\ket$ such that $|e_{m,j}-e_{n,i} -\omega| \leq \delta$, we can extend the above procedure and express the resulting sum in terms of the following operator,
 \be
 {\cal A}(\omega):= \sum_{m,n,i,j;|e_{m,j}-e_{n,i} -\omega|\leq \delta}A_{n,i,m,j}|n,i\ket\bra m,j|.
 \ee
 With such operator, for all times $t$ such that $(g^2\tau_c)^{-1}\leq t\ll \delta^{-1}$, the above master equation \eqref{derme} can be rewritten as,
 \bea\label{mebis}
 \dot \rho_t = \sum_{\omega} \Gamma(\omega) \big[{\cal A}(\omega)\rho_t{\cal A}^{\dag}(\omega) -{\cal A}^{\dag})\omega) {\cal A}(\omega) \rho_t\big] + {\rm h.c.},\nn\\
 \eea
   which coincides with the form of the master equation \eqref{appme}. Note that the condition $t \geq (g^2\tau_c)^{-1}$ is the usual condition for the validity of the secular approximation. The new condition here is $t\ll\delta^{-1}$, which requires $\delta^{-1}\gg(g^2\tau_c)^{-1}$ due to the previous condition. Since the decay rates are of the order $|\Gamma(\omega)|\sim g^2\tau_c$, we can conclude that the order of magnitude of the inhomogeneities or interactions has to be smaller than the decay rate in oder to have a reduced dynamics of the form \eqref{me}. This regime can be achieved for instance with atomic ensembles in an optical cavity in the limit of bad cavity \cite{Barberena_2019}, already realisable in some experimental platforms \cite{Norcia_2016,Norcia_2018}. 
   This also means that the bath does not resolve energy differences of order $\delta$ until $ t\sim \delta^{-1}$. In other words, the diverse effects mentioned in the main text can go on in principle over a period of time of the order of $\delta^{-1}$. \\

\section{Time derivatives of ${\cal C}_{\rm v}$, ${\cal C}_{\rm h}$, and ${\cal D}_{\rm th}$}  \label{apptimeder}
While it is well-known that the time derivative of the von Neumann entropy is $-{\rm Tr}\dot \rho_t \ln\rho_t$, it is not straightforward that similar relations hold for ${\cal C}_{\rm v}$, ${\cal C}_{\rm h}$, and ${\cal D}_{\rm th}$. In order to prove the identities (4) of the main text we only need to show 
\be\label{tderdiag}
\frac{ d}{dt} {\rm Tr} \rho_t \ln \rho_{t|_{\rm D}} = {\rm Tr} \dot \rho_t \ln \rho_{t|_{\rm D}},
\ee
and,
\be\label{tderbd}
\frac{ d}{dt} {\rm Tr} \rho_t \ln \rho_{t|_{\rm BD}} = {\rm Tr} \dot \rho_t \ln \rho_{t|_{\rm BD}}.
\ee
Starting with the first equality \eqref{tderdiag}, from the definition of $\rho_{t|_{\rm D}}$ 
we have $\rho_{t|_{\rm D}} = \sum_{n}\sum_{i=1}^{l_n}p_{n,i}|n,i\ket\bra n,i|$, where $p_{n,i}:= \bra n,i|\rho_t|n,i\ket$. Thus, $\ln \rho_{t|_{\rm D}} = \sum_{n}\sum_{i=1}^{l_n}|n,i\ket\bra n,i|\ln p_{n,i}$ so that 
  \be
{\rm Tr} \rho_t \ln \rho_{t|_{\rm D}}=\sum_{n}\sum_{i=1}^{l_n}p_{n,i}\ln p_{n,i}.
 \ee
 Note that the above equations shows 
 \be\label{identity1}
 {\rm Tr} \rho_t \ln \rho_{t|_{\rm D}}={\rm Tr} \rho_{t|_{\rm D}} \ln \rho_{t|_{\rm D}}.
 \ee
Denoting by $\dot p_{n,i}:=\bra n,i|\dot \rho_t|n,i\ket$ the time derivative of the populations $p_{n,i}$, one obtains
\bea
\frac{ d}{dt} {\rm Tr} \rho_t \ln \rho_{t|_{\rm D}} &=& \sum_{n}\sum_{i=1}^{l_n}\dot p_{n,i}\ln p_{n,i} +  \sum_{n}\sum_{i=1}^{l_n}\dot p_{n,i}\nn\\
&=& {\rm Tr} \dot \rho_t \ln \rho_{t|_{\rm D}}
\eea
since $\sum_{n}\sum_{i=1}^{l_n}\dot p_{n,i}={\rm Tr} \dot \rho_t =0$.

For the second equality \eqref{tderbd}, we can proceed in a similar way. Using the definition of $\rho_{t|_{\rm BD}}$ 
we have 
\bea
\rho_{t|_{\rm BD}} &=& \sum_{n}\pi_n \rho_t \pi_n\nn\\
&=& \sum_{n}\sum_{i,i'}|n,i\ket\bra n,i| \rho_t |n,i'\ket\bra n,i'|.
\eea
which is diagonal per block (corresponding to each eigenspace). Denoting by $\{|e_{n,i}\ket\}_{1\leq i\leq l_n}$ a basis diagonalising $\rho_{t|_{\rm BD}}$ on each eigenspace we have $\rho_{t|_{\rm BD}} = \sum_{n}\sum_{i=1}^{l_n}q_{n,i}|e_{n,i}\ket\bra e_{n,i}|$, where $q_{n,i}:= \bra e_{n,i}|\rho_t|e_{n,i}\ket$. It is important to keep in mind that first, this basis is time-dependent, and secondly, due to the block-diagonal structure of $\rho_{t|_{\rm BD}} $, each $|e_{n,i}\ket$ is a linear combination of exclusively the eigenvectors $\{|n,i\ket\}_{1\leq i\leq l_n}$ (spanning the eigenspace $n$). This two observations will be crucial in the following.
 As previously, we have
 \be
{\rm Tr} \rho_t \ln \rho_{t|_{\rm BD}}=\sum_{n}\sum_{i=1}^{l_n}q_{n,i}\ln q_{n,i},
 \ee
which also shows 
\be\label{identity2}
{\rm Tr} \rho_t \ln \rho_{t|_{\rm BD}}={\rm Tr} \rho_{t|_{\rm BD}} \ln \rho_{t|_{\rm BD}}.
\ee 
Differently from the previous situation, the diagonalisation basis $\{|e_{n,j}\ket\}_{1\leq j\leq l_n}$ is time dependent. This implies
\be\label{derqni}
\dot q_{n,i} = \frac{d}{dt}(\bra e_{n,i}|) \rho_t |e_{n,i}\ket + \bra e_{n,i}| \dot\rho_t |e_{n,i}\ket+\bra e_{n,i}| \rho_t\frac{d}{dt} (|e_{n,i}\ket).
\ee
Then,
\be
\frac{d}{dt} {\rm Tr} \rho_t \ln \rho_{t|_{\rm BD}}=\sum_{n}\sum_{i=1}^{l_n}[\dot q_{n,i}\ln q_{n,i} + \dot q_{n,i}],
\ee
where $\dot q_{n,i}$ is given by the above expression \eqref{derqni}. We have,
\bea
\sum_{n}\sum_{i=1}^{l_n} \dot q_{n,i}& =& \frac{d}{dt}\sum_{n}\sum_{i=1}^{l_n} \bra e_{n,i}|\rho_t|e_{n,i}\ket\nn\\
&=& \frac{d}{dt} {\rm Tr} \rho_t\nn\\
&=& 0
\eea
The second term is slightly more involved,
\bea
\sum_{n}\sum_{i=1}^{l_n}\dot q_{n,i}\ln q_{n,i} &=&\sum_{n}\sum_{i=1}^{l_n}\bra e_{n,i}| \dot\rho_t |e_{n,i}\ket\ln q_{n,i} \nn\\
&&+\sum_{n}\sum_{i=1}^{l_n} \frac{d}{dt}(\bra e_{n,i}|) \rho_t  |e_{n,i}\ket\ln q_{n,i} \nn\\
&&+\sum_{n}\sum_{i=1}^{l_n} \ln q_{n,i} \bra e_{n,i}| \rho_t \frac{d}{dt} (|e_{n,i}\ket).\nn\\
\eea
 Using the identity $|e_{n,i}\ket\ln q_{n,i}=\ln \rho_{t|_{\rm BD}} |e_{n,i}\ket$ the first term is equal to 
\be
\sum_{n}\sum_{i=1}^{l_n}\bra e_{n,i}| \dot\rho_t |e_{n,i}\ket\ln q_{n,i}={\rm Tr} \dot \rho_t \ln  \rho_{t|_{\rm BD}}. 
\ee
The last two terms sum up to zero. This can be shown for instance by introducing the decomposition of the identity $1=\sum_{n}\pi_n$ on both sides of $\rho_t$ in each term. Importantly, as mention previously, since each vector $|e_{n,i}\ket$ is a linear combination of exclusively the eigenvectors $\{|n,i\ket\}_{1\leq i\leq l_n}$, its time derivative $\frac{d}{dt}|e_{n,i}\ket$ belongs to the eigenspace $n$ (spanned by the vectors $\{|n,i\ket\}_{1\leq i\leq l_n}$). This implies in particular that $\pi_m\frac{d}{dt}|e_{n,i}\ket =0$ if $n\ne m$. 
%
%
%
%
%
%
%
We obtain, 
\bea
&&\sum_{n}\sum_{i=1}^{l_n} \frac{d}{dt}(\bra e_{n,i}|) \rho_t  |e_{n,i}\ket\ln q_{n,i} \nn\\
&&=\sum_{n}\sum_{i=1}^{l_n} \frac{d}{dt}(\bra e_{n,i}|) \sum_m\pi_m\rho_t \sum_{m'}\pi_{m'}  |e_{n,i}\ket\ln q_{n,i} \nn\\
&&=\sum_{n}\sum_{i=1}^{l_n} \frac{d}{dt}(\bra e_{n,i}|) \pi_n\rho_t \pi_{n}  |e_{n,i}\ket\ln q_{n,i}\nn\\
&&=\sum_{n}\sum_{i=1}^{l_n} \frac{d}{dt}(\bra e_{n,i}|) \rho_{t|_{\rm BD}}  |e_{n,i}\ket\ln q_{n,i}\nn\\
&&=\sum_{n}\sum_{i=1}^{l_n} \frac{d}{dt}(\bra e_{n,i}|)  |e_{n,i}\ket q_{n,i}\ln q_{n,i}.
\eea
Proceeding in a similarly for the term $\sum_{n}\sum_{i=1}^{l_n} \ln q_{n,i} \bra e_{n,i}| \rho_t \frac{d}{dt} (|e_{n,i}\ket)$, one finally obtains
\bea
&&+\sum_{n}\sum_{i=1}^{l_n} \frac{d}{dt}(\bra e_{n,i}|) \rho_t  |e_{n,i}\ket\ln q_{n,i} \nn\\
&&+\sum_{n}\sum_{i=1}^{l_n} \ln q_{n,i} \bra e_{n,i}| \rho_t \frac{d}{dt} (|e_{n,i}\ket)\nn\\
&&=\sum_{n}\sum_{i=1}^{l_n}\left[ \frac{d}{dt}(\bra e_{n,i}|)  |e_{n,i}\ket  + \bra e_{n,i}|\frac{d}{dt} (|e_{n,i}\ket) \right]q_{n,i}\ln q_{n,i}\nn\\
&&=\sum_{n}\sum_{i=1}^{l_n}q_{n,i}\ln q_{n,i}\frac{d}{dt}\left[ \bra e_{n,i}|  e_{n,i}\ket \right]\nn\\
&&=0.
\eea
Then, all together we have shown what we announced above, equations \eqref{tderdiag} and \eqref{tderbd}, completing the demonstration of the identities (4) in the main text.\\

\section{Collective coupling and indistinguishability}\label{appcolcoupling}
%
 The most general coupling between $S$ and $B$ is of the form
\be
V=g \sum_{\alpha} A_{S,\alpha} A_{B,\alpha},
\ee
where $g$ corresponds to the effective coupling strength, $A_{S,\alpha}$ observables of $S$ and $A_{B,\alpha}$ observables of the bath. When two different bath observables $A_{B,\alpha}$ and $A_{B,\alpha'}$ are independent such that ${\rm Tr}\rho_B A_{B,\alpha}A_{B,\alpha'} =0$, each observable give rise to an {\it independent dissipation} channel, as if $A_{B,\alpha}$ and $A_{B,\alpha'}$ were observables of two distinct and independent baths. Contrasting with such {\it independent dissipation}, we assume in this paper a situation where the system-bath coupling give rise to a single dissipation channel only, which corresponds to a coupling of the alternative following form
\be\label{colcou}
V=g  A_{S}A_B.
\ee
It means that all energy transitions are {\it collectively} coupled to the {\it same bath observable} $A_B$. 
In particular, an absorption of a bath excitation can activate any resonant transition, ending up in any corresponding excited state. Thus, the bath does not ``know'' which transition was activated: it cannot distinguish two (or several) different resonant transitions. 
This interpretation provides some insights regarding the underlying conditions for collective coupling \eqref{colcou}, namely, the transitions should be indistinguishable from the point of view of the bath.   
Depending on the system, this might require some experimental arrangements. For instance for a multi-level atom, one needs parallel transition dipole moments \cite{Tscherbul_2015}, realised experimentally in \cite{Xia_1996}, or optical cavity to make the atomic transitions indistinguishable (from the point of view of the outside bath) \cite{Patnaik_1999,Zhou_2001}. Conversely, if $S$ is made of an ensemble of smaller subsystems, each subsystem should be placed at spatial locations which are indistinguishable, or indiscernible, from the point of view of bath \cite{Gross_1982,bathinducedcohTLS} (with an example of experimental realisations in \cite{Devoe_1996, Krauter_2011}). 
Alternatively, the indistinguishability from the bath's point of view can be achieved also by adding an ancillary system between $S$ and the bath, like an optical cavity in the situations of atomic clouds \cite{Wood_2014, Wood_2016, Hama_2018, Niedenzu_2018}, with examples of experimental realisations in \cite{Raimond_1982, Barberena_2019}.

The reduced dynamics of $S$ can be obtained using the Born and Markov approximations, valid for weak bath coupling \cite{Cohen_Book,Petruccione_Book}, leading to the following master equation (Eq. \eqref{me} of the main text) thanks to the additional secular approximation,
\bea\label{appme}
\dot{\rho}_t &=& {\cal L} \rho_t \nn\\
&:=&  \sum_{\omega} \Gamma(\omega) \left[{\cal A}(\omega)\rho_t {\cal A}^{\dag}(\omega) -{\cal A}^{\dag}(\omega){\cal A}(\omega)\rho_t\right] + {\rm h.c.},\nn\\
\eea
recalling the notations $\Gamma(\omega):=\int_0^{\infty}ds e^{i\omega s}{\rm Tr}\rho_B A_B(s)A_B$, $A_B(s)$ is the bath operator $A_B$ in the interaction picture (with respect to the free Hamiltonian $H_B$), and the jump operators ${\cal A}(\omega)$ are defined by \cite{Petruccione_Book} ${\cal A}(\omega)=\sum_{e_{n'}-e_n=\omega}  \pi_n A_S \pi_{n'}$.
 We now emphasise the structure imposed by the master equation \eqref{appme} on the dynamics of the populations, vertical coherences, and horizontal coherences. From \eqref{appme}, the dynamics of the matrix element $\bra n,i|\rho_S|n',i'\ket$ is
\bea\label{coherencedynamics}
 \frac{d}{dt}\bra n,i|\rho_S|n',i'\ket &=& \sum_{\omega} \Gamma(\omega) \Big[{\bra n,i|\cal A}(\omega)\rho_t {\cal A}^{\dag}(\omega)|n',i'\ket \nn\\
 &&-\bra n,i|{\cal A}^{\dag}(\omega){\cal A}(\omega)\rho_t|n',i'\ket\Big] + {\rm c.c.}\nn\\
\eea
Then, if $n=n'$, the states ${\cal A}^{\dag}(\omega)|n',i'\ket$ and ${\cal A}(\omega)|n,i\ket$ lay in the eigenspace of energy $e_{n'}+\omega=e_n +\omega$, so that the terms $\bra n,i|{\cal A}(\omega)\rho_t {\cal A}^{\dag}(\omega)|n',i'\ket$ corresponds to horizontal coherences and populations. The state ${\cal A}^{\dag}(\omega){\cal A}(\omega)|n,i\ket$ belongs to the eigenspace of energy $e_n$ so that $\bra n,i|{\cal A}^{\dag}(\omega){\cal A}(\omega)\rho_t|n',i'\ket$ also corresponds to horizontal coherences or populations (still when $n=n'$). In other words, the dynamics of the populations and horizontal coherences are coupled. By contrast, if $n\ne n'$, one can see that only vertical coherences appear on the the right-hand side of \eqref{coherencedynamics}, so that the dynamics of vertical coherences is not coupled neither to the populations nor to the horizontal coherences.\\

\section{Entropy production in non-degenerate systems}\label{appnondegen}
 For non-degenerate systems, $\rho_{t|_{\rm D}}=\rho_{t|_{\rm BD}}$, so that ${\cal C}_{\rm h}=0$ at all times. Moreover, $-\dot {\cal C}_{\rm v}$ and $-\dot {\cal D}_{\rm th}$ become equivalent to the coherent and diagonal contributions introduced in Eq. (12) of \cite{Santos_2019}. One can also show that this two remaining contributions to the entropy production are always positive. Considering the time derivative of the relative entropy of coherences one obtains
\bea\label{positivitycen}
-\dot {\cal C}_{\rm v} &=& -\frac{d}{dt} S(\rho_t|\rho_{t|_{\rm BD}})= -\frac{d}{dt} S(\rho_t|\rho_{t|_{\rm D}})\nn\\
 &=&-\lim_{dt\rightarrow 0} \frac{1}{dt} \Big[S(\rho_{t+dt}|\rho_{t+dt|_{\rm D}})-S(\rho_{t}|\rho_{t|_{\rm D}})\Big]\nn\\
&=&-\lim_{dt\rightarrow 0} \frac{1}{dt} \Big[S(e^{dt {\cal L}}\rho_{t}|e^{dt{\cal L}}\rho_{t|_{\rm D}})-S(\rho_{t}|\rho_{t|_{\rm D}})\Big],\nn\\
\eea
which is always positive since the relative entropy is contractive under completely positive and trace preserving maps \cite{Lindblad_1975,Petruccione_Book} (the map generated by ${\cal L}$ defined in \eqref{appme} being completely positive and trace preserving). Similarly, one can show that velocity of the population convergence to the thermal equilibrium distribution $-\dot {\cal D}_{\rm th}$ is, as expected, always positive, recovering the results of \cite{Santos_2019}.
The above result relies on the following crucial step,
\be\label{equality}
 \rho_{t+dt|_{\rm D}} = e^{dt {\cal L}}\rho_{t|_{\rm D}}, 
 \ee
 which means that the dynamics of the populations depends only on the populations themselves (remembering that we are considering {\it non-degenerate} systems). In other words, the presence of coherences does not influence the future values of the populations. This is a fundamental difference with degenerate systems, as shown in the main text. Equation \eqref{equality} is also equivalent to the Pauli equation \cite{Petruccione_Book, Santos_2019} (which gives the dynamics of the populations in terms of themselves), and to \eqref{coherencedynamics} applied to non-degenerate systems.\\

\section{Reduction of irreversibility in spin ensembles} \label{appss}
In this paragraph we recall some properties of spin ensembles and theory of addition of angular momentum. We also recall the expression of the equilibrium state reached by a spin ensemble when collective interacting with a bath (as described for instance by Eq. \eqref{me} of the main text).
Considering an ensemble containing $n$ spins of size $s$, we denote by $j_{z,k}$ the z-component of the angular momentum operator associated to the spin $k \in [1;n]$, and by $\{|s,m_k\ket_k \}_{-s\leq m_k\leq s}$ the local eigenbasis of $j_{z,k}$, so that $j_{z,k}|s,m_k\ket_k = \hbar m_k |s,m_k\ket_k$. Then, a natural basis to describe the spin ensemble is 
 \be\label{localbasis}
|m_1,m_2,...,m_n\ket :=\otimes_{k=1}^n |s,m_k\ket_k,
\ee
 resulting from the tensor products of the local eigenbasis. 
 One important property from the theory of addition of angular momenta \cite{Sakurai_Book} is that the spin ensemble can be described alternatively by a basis obtained from the eigenvectors of the global observables $J_z$ and ${\cal J}^2:=J_x^2+J_y^2+J_z^2$, where $J_z:=\sum_{k=1}^n j_{z,k}$ (and similar definitions for the $x$ and $y$ components). These eigenvectors are traditionally denoted by $|J,m\ket_i$ in reference to their eigenvalues, 
 \bea
&&{\cal J}^2|J,m\ket_i = \hbar J(J+1)|J,m\ket_i\nn\\
&& J_z|J,m\ket_i= \hbar m|J,m\ket_i,
\eea
 with $-J\leq m\leq J$ and $J \in [J_0; ns]$, where $J_0 =0$ if $s\geq 1$ and $J_0=1/2$ if $s=1/2$ and $n$ odd. The index $i$ belongs to the interval $[1;l_J]$, where $l_J$ denotes the degeneracy of the eigenspace associated to the eigenvalue $J$ of the total spin operator ${\cal J}^2$.

The equilibrium state reached by the spin ensemble initially in a thermal state at inverse temperature $\beta_0$ and interacting collectively with a bath at inverse temperature $\beta_B$ is given by \cite{bathinducedcoh}
\be\label{steadystate}
\rho^{\infty}_{\beta_0}(\beta_B) := \sum_{J=J_0}^{ns}p_J(\beta_0)\sum_{i=1}^{l_J}\rho_{J,i}^{\rm th}(\beta_B),
\ee
where $p_J(\beta_0):= Z_J(\beta_0)/Z(\beta_0)$, $Z_J(\beta_0):= \sum_{m=-J}^J e^{-m\hbar \omega \beta_0}$, $Z(\beta_0):=(Z_s(\beta_0))^n$, and $\rho_{J,i}^{\rm th}(\beta_B):=Z_{J}(\beta_B)^{-1}\sum_{m=-J}^J e^{- m\hbar\omega\beta_B}|J,m\ket_i\bra J,m|$.

The details of the expression \eqref{steadystate} are not essential here. What is important however is that $\rho^{\infty}_{\beta_0}(\beta_B)$ contains horizontal coherences (in the natural basis $|m_1,m_2,...,m_n\ket$) whenever $\beta_0 \ne \pm \beta_B$. This can be seen by the following considerations.
First, note that there is a unique equilibrium state which is diagonal (in the natural basis). The reason is because any diagonal equilibrium state should satisfy the detailed balance, but there is only one diagonal state satisfying it: $\rho^{\rm th}(\beta_B)$, the thermal state of inverse temperature $\beta_B$.
 This thermal state is reached for $\beta_0=\pm\beta_B$. When $\beta_0 \ne \pm \beta_B$, the energy of $\rho^{\infty}_{\beta_0}(\beta_B)$ is different from the thermal energy of $\rho^{\rm th}(\beta_B)$ \cite{bathinducedcoh}. Therefore, for any $\beta_0 \ne \pm \beta_B$, the equilibrium state  cannot be equal to $\rho^{\rm th}(\beta_B)$ and consequently cannot be a diagonal state. Moreover,  $\rho^{\infty}_{\beta_0}(\beta_B)$ does not contain any vertical coherences since it is made up of statistical mixtures of collective spin states $|J,m\ket_i\bra J,m|$, themselves containing no vertical coherences, $\bra m_1,...,m_n|J,m\ket_i\bra J,m|m_1',...m_n'\ket=0$ if $m_1+...+m_n\ne m_1'+...+m_n'$. Consequently, $\rho^{\infty}_{\beta_0}(\beta_B)$ necessarily contains horizontal coherences, implying $\rho^{\infty}_{\beta_0}(\beta_B)_{|_{\rm D}}\ne\rho^{\infty}_{\beta_0}(\beta_B)_{|_{\rm BD}}=\rho^{\infty}_{\beta_0}(\beta_B)$ so that $-{\cal C}_{\rm h}(\infty) = - S[\rho^{\infty}_{\beta_0}(\beta_B)_{|_{\rm BD}}|\rho^{\infty}_{\beta_0}(\beta_B)_{|_{\rm D}}] <0$.
 
 Thus, since the ensemble is initially in a thermal state (with zero horizontal coherences), we have creation of horizontal coherences which corresponds to $-\Delta^{\infty} {\cal C}_{\rm h} = -[{\cal C}_{\rm h}(\infty)- {\cal C}_{\rm h}(0)]=-{\cal C}_{\rm h}(\infty)< 0$, promoting a reduction of the entropy production. 
 We can even show that the value of $\Delta^{\infty} {\cal C}_{\rm h}$ increases logarithmically with the number of degenerate levels (growing with $n$ and $s$), at least in the limit $\omega |\beta_0|\gg1$. 
From \eqref{steadystate} one can obtain the following expression for the diagonal cut \cite{bathinducedcoh},
\bea\label{diagonalcut}
\rho^{\infty}_{\beta_0}(\beta_B)_{|_{\rm D}}&\underset{\hbar\omega|\beta_0|\gg1}{=}&Z_{ns}^{-1}(\beta_B) \sum_{m=-ns}^{ns} ~\sum_{m_1+...+m_n=m} \nn\\
&&\frac{e^{-\hbar\omega m\beta_B}}{I_m}|m_1,...,m_n\ket \bra m_1,...,m_n|,\nn\\
\eea
where $I_m:=\sum_{J=|m|}^{ns} l_J$ is the total number of eigenstates of $J_z$ of eigenvalue $\hbar m$. One can deduce from \eqref{diagonalcut} the expression for the variation of horizontal coherences,
\be
-\Delta^{\infty} {\cal C}_{\rm h}\!=\!-{\cal C}_{\rm h}(\infty) \!\underset{\hbar\omega|\beta_0|\gg1}{=}\!\! -\!\sum_{m=-ns}^{ns} \frac{e^{-\hbar\omega m\beta_B}}{Z_{ns}(\beta_B)} \ln I_m,
\ee
so that $\Delta^{\infty} {\cal C}_{\rm h}$ grows logarithmically with the number of degenerate levels $I_m$ (itself a growing function of $n$ and $S$).

Beyond that, the variation of $-{\cal D}_{\rm th}$, $-\Delta^{\infty} {\cal D}_{\rm th} :=-[{\cal D}_{\rm th}(\infty)-{\cal D}_{\rm th}(0)]$, is also reduced (compared to independent dissipation).
 This is can be seen simply as follows. Since the equilibrium state energy is different from the thermal equilibrium energy \cite{bathinducedcoh}, we have necessarily $\rho^{\infty}_{\beta_0}(\beta_B)_{|_{\rm D}}\ne\rho^{\rm th}(\beta_B)$. Consequently, the measure of distance to the thermal distribution $\rho^{\rm th}(\beta_B)$ does not tend to zero but to a strictly positive value, ${\cal D}_{\rm th}(\infty)_{|\rm col} = S[\rho^{\infty}_{\beta_0}(\beta_B)_{|_{\rm D}}|\rho^{\rm th}(\beta_B)] >0$. By comparison, for independent dissipation, the ensemble reaches the equilibrium thermal state $\rho^{\rm th}(\beta_B)$ so that ${\cal D}_{\rm th}(\infty)_{|\rm ind} =0$.
 This makes the variation $ -\Delta^{\infty} {\cal D}_{\rm th}$ strictly smaller for collective dissipation than for independent dissipation. \\

\section{Expression of the heat flow} \label{apphflow}
The heat flow between the system and the bath is defined by \cite{Alicki_1979} 
\be
\dot E_S := {\rm Tr} \dot \rho_t H_S.
\ee
 Using the dynamics described by \eqref{appme} as the time derivative of $\rho_t$, one obtains
 \bea
 \dot E_S &=& \sum_{\omega} \Gamma(\omega) {\rm Tr} \Big(\ca\rho_t\cad -\cad\ca\rho_t\Big)H_S \nn\\
 &&+ {\rm c.c.}\nn\\
 &=& \sum_{\omega} \Gamma(\omega) {\rm Tr} \ca\rho_t[\cad,H_S] + {\rm c.c.}\nn\\
  &=& -\sum_{\omega} \omega\Gamma(\omega) {\rm Tr} \ca\rho_t\cad + {\rm c.c.}\nn\\
  &=& -\sum_{\omega} \omega G(\omega) \bra\cad\ca\ket_{\rho_t}, \nn\\
 \eea
where $\bra {\cal O}\ket_{\rho_t} := {\rm Tr}\rho_t {\cal O}$ stands for the expectation value of any operator ${\cal O}$ taken in the state $\rho_t$. Furthermore, we used in the second line the invariance of the trace under cyclic permutations, in the third line, the commutation relation of the eigenoperators $[\cad,H_S]=-\omega\cad$,  and in the fourth line, the definition $G(\omega):= \Gamma(\omega)+\Gamma^{*}(\omega)$ already introduced previously. 
We can rewrite the above expression in an insightful way by explicitly including the negative frequencies,
\bea
\dot E_S &=& -\sum_{\omega>0} \omega \Big(G(\omega) \bra\cad\ca\ket_{\rho_t} \nn\\
 &&\hspace{2cm}-G(-\omega) \bra\ca\cad\ket_{\rho_t}\Big) \nn\\
 &=&\sum_{\omega>0}\omega G(\omega)\bra\ca\cad\ket_{\rho_t}\big(e^{-\omega\beta_B}-e^{-\omega/{\cal T}(\omega)}\big),\nn\\
 \eea
where ${\cal T}(\omega):= \omega \left(\ln\frac{\bra \ca\cad\ket_{\rho_t}}{\bra\cad\ca\ket_{\rho_t}}\right)^{-1}$ is the apparent temperature associated with the energy exchange $\omega$ \cite{paperapptemp,autonomousmachines}, and the bath inverse temperature $\beta_B$ can be defined through $e^{-\beta\omega}:= G(-\omega)/G(\omega)$ \cite{paperapptemp,Alicki_2014,Alicki_2015}. In particular, when the population of the state $\rho_t$ follows a thermal distribution, as for $\rho_0$  in Eq. (15) of the main text, one can express the inverse apparent temperature as 
\be
\frac{\omega}{{\cal T}(\omega)} = \omega\beta_0 +  \ln \frac{1+ c^{+}}{1+c^{-}},
\ee
where $\beta_0$ is the inverse temperature associated with the thermal distribution of the populations, and $c^{-}:= \bra \ca^{\dag} \ca \ket_{\chi}/\bra \ca^{\dag} \ca \ket_{\rho^{\rm th}(\beta_0)}$ and $c^{+}:=\bra \ca \ca^{\dag} \ket_{\chi}/\bra \ca \ca^{\dag} \ket_{\rho^{\rm th}(\beta_0)}$ 
constitute the contributions from the {\it horizontal coherences}, highlighted in \cite{paperapptemp,heatflowreversal,autonomousmachines}. Indeed, since $\ca\cad$ and $\cad\ca$ commute with $H_S$, their expectation values do not pick up contributions from vertical coherences. In other words, when $\chi$ do not contain {\it horizontal coherences} one has $\bra\ca\cad\ket_{\chi}=\bra\cad\ca\ket_{\chi}=0$ implying ${\cal T}(\omega)=1/\beta_0$. \\

\section{Thermal operations and beyond}\label{appbroad}
{\it Positivity of $-\Delta {\cal C}_{\rm v}$ and positivity break down for ${\cal C}_{\rm h}$ and ${\cal D}_{\rm th}$}. 
In this section we show that the simple conditions of energy conservation, initial separability, and initial stationarity of $B$ ($[H_B,\rho_{B,t_i}]=0$) have rich consequences. As introduced in the main text, we consider that the systems $S$ and $B$ interact unitarily through $U$ from $t_i$ to $t_f$. The reduced dynamics for $S$ is given by,
\bea
\rho_{S, t_f}= \Lambda \rho_{S, t_i} &:=& {\rm Tr}_B U\rho_{SB, t_i} U^{\dag}\nn\\
&=& \sum_{\nu,\mu} M_{\mu,\nu} \rho_{S,t_i} M_{\mu,\nu}^{\dag}
\eea
where $M_{\mu,\nu}:= \sqrt{p_\nu} \bra \psi_\mu|U|\psi_\nu\ket$, $p_\nu:=\bra \psi_\nu|\rho_{B,t_i}|\psi_\nu\ket$, and the eigenstates and eigenenergies of $B$ are denoted respectively by $|\psi_\nu\ket$ and $E_\nu$. We also denote by ${\cal H}_{S,n}$ the eigenspace of $S$ associated with the energy $e_n$. Since the initial state of $B$ is assumed to be stationary it can be written in the form $\rho_{B,t_i}=\sum_{\nu} p_\nu |\psi_\nu\ket\bra\psi_\nu|$. Note that we include the possibility of $B$ being degenerate, but in order to simplify the notation we do not explicit write an extra index representing the degeneracy. It means that several $E_\nu$ and $p_\nu$ can have the same value. The system $B$ can also have an unbounded discrete spectrum or even a continuous spectrum (if we think of an infinite bath). In this later situation one should express $H_B$ and $\rho_B$ through integrals. For simplicity again we maintain the discrete sum notations but one should bear in mind that the following results can be extended to continuous spectrum.  
From the above notations, the final (at time $t_f$) vertical coherences between the state $|n,i\ket $ and $|m,j\ket$ can be expressed as,
\bea\label{d2}
\bra n,i|\rho_{S,t_f}|m,j\ket &=& \sum_{\mu,\nu} \bra n,i| M_{\mu,\nu} \rho_{S, t_i} M_{\mu,\nu}^{\dag}  |m,j\ket\nn\\
&=&\sum_{\mu,\nu,q,r} \bra n,i| M_{\mu,\nu}\pi_q \rho_{S, t_i}\pi_r M_{\mu,\nu}^{\dag}  |m,j\ket.\nn\\
\eea 
``Sandwiching'' the conservation energy relation $[U,H_S+H_B]=0$ between $\bra n,i|\bra\psi_\mu|$ and $|q,l\ket|\psi_\nu\ket$ we obtain
\be
\bra n,i|M_{\mu,\nu} |q,l\ket (e_q +E_\nu -e_n -E_\mu)=0.
\ee
This implies in \eqref{d2} that if $q$ is such that $e_q \ne e_n +E_\mu-E_\nu$, we must have $ \bra n,i| M_{\mu,\nu}\pi_q=0$. Similarly for the term $\pi_r M_{\mu,\nu}^{\dag}  |m,j\ket$. One may conclude that the only term contributing to \eqref{d2} are such that $e_q = e_n +E_\mu-E_\nu$ and $e_r = e_m +E_\mu-E_\nu$ so that if $e_n\ne e_m$, we have necessarily $e_q\ne e_r$. Consequently, only vertical coherences contribute to the sum \eqref{d2}: the dynamics of vertical coherences depends only on vertical coherences. 
By contrast, if $e_n=e_m$, we have necessarily $e_q=e_r$ (and finally $q=r$), so that only horizontal coherences and populations contribute to \eqref{d2}. In other words, the dynamics of the populations and horizontal coherences are coupled, but both are {\it decoupled} from the dynamics of the vertical coherences. Consequently, the following identity holds,
\be\label{appbdid}
\rho_{S,t_f|_{\rm BD} }= \Lambda \rho_{S,t_i|_{\rm BD}},
\ee
whereas
\be
\rho_{S,t_f|_{\rm D} }\ne \Lambda \rho_{S,t_i|_{\rm D}}.
\ee
Then, one can repeat the argument in Eq. (6) of the main text to show that the variation of $-{\cal C}_{\rm v}(t)$ between $t_i$ and $t_f$ is always positive (meaning that vertical coherences are consumed). Namely,
\bea
-\Delta {\cal C}_{\rm v} &=& -[{\cal C}_{\rm v}(t_f)-{\cal C}_{\rm v}(t_i)]\nn\\
&=& -[S(\rho_{S,t_f}|\rho_{S,t_f|_{\rm BD}})-S(\rho_{S,t_i}|\rho_{S,t_i|_{\rm BD}})]\nn\\
&=& -[S(\Lambda\rho_{S,t_i}|\Lambda\rho_{S,t_i|_{\rm BD}})-S(\rho_{S,t_i}|\rho_{S,t_i|_{\rm BD}})] \geq 0,\nn\\
\eea
where the positivity comes from the contractivity of the relative entropy under completely positive and trace preserving maps (of which $\Lambda$ belongs since we assume $S$ and $B$ initially uncorrelated \cite{Petruccione_Book}). 
By contrast, since $\rho_{S,t_f|_{\rm D} }\ne \Lambda \rho_{S,t_i|_{\rm D}}$, the guarantee of positivity breaks down for $-\Delta {\cal C}_{\rm h}$ and $-\Delta {\cal D}_{\rm th}$, where ${\cal D}_{\rm th}(t)$ is defined assuming the existence of a thermal equilibrium state $\rho_{S}^{\rm th}(\beta_B)$. Note that this is guaranteed at least when $B$ is initially in a thermal state at inverse temperature $\beta_B$ (thanks to energy conservation, $S$ always admits the thermal state $\rho_S^{\rm th}(\beta_B)$ as equilibrium state ). In such situation $U$ corresponds to the well-known thermal operations \cite{Janzing_2000}.

On the other hand, the sum of  $-\Delta {\cal C}_{\rm h}$ and $-\Delta {\cal D}_{\rm th}$ always remains positive,
\bea
&&-\Delta {\cal C}_{\rm h} \!\!-\!\Delta {\cal D}_{\rm th} \!= \!-[{\cal C}_{\rm h}(t_f)\!+\!{\cal D}_{\rm th}(t_f)\!-\!{\cal C}_{\rm h}(t_i)\!-\!{\cal D}_{\rm th}(t_i)]\nn\\
&&\hspace{1cm}=\! -\{S[\rho_{S,t_f|_{\rm BD}}|\rho_S^{\rm th}(\beta_B)]-S[\rho_{S,t_i|_{\rm BD}}|\rho_S^{\rm th}(\beta_B)]\}\nn\\
&&\hspace{1cm}=\!-\{S[\Lambda\rho_{S,t_i|_{\rm BD}}|\Lambda\rho_S^{\rm th}(\beta_B)]-S[\rho_{S,t_i|_{\rm BD}}|\rho_S^{\rm th}(\beta_B)]\}\nn\\
&&\hspace{1cm} \geq 0,
\eea
as announced in the main text.

 Similarly to what have been done in the main text for the bath-driven dissipation, one can pinpoint explicit situations where $-\Delta {\cal C}_{\rm h} <0$. 
 Denoting by $|\psi_\nu\ket$ and $E_{\nu}$ the (possibly degenerate) eigenstates and eigenenergies of $B$, one consequence of the energy conservation is that the transition $\bra n,i|\bra \psi_\nu|U|m,j\ket|\psi_\mu\ket$ is equal to zero unless $e_n+E_{\nu}-e_{m}-E_{\mu}=0$. In particular, one can have degenerate transitions from one state $|m,j\ket|\psi_\mu\ket$ to two degenerate states $|n,i\ket|\psi_\nu\ket$ and $|n,i'\ket|\psi_\nu\ket$, expressed by $\bra n,i|\bra \psi_\nu|U|m,j\ket |\psi_\mu\ket\ne 0$ and $\bra n,i'|\bra \psi_\nu|U|m,j\ket|\psi_\mu\ket\ne 0$. Then, if for instance $SB$ is initialised in the state $|m,j\ket|\psi_\mu\ket$, such a unitary evolution $U$ definitively generates horizontal coherences and a negative contribution to the entropy production, $-\Delta {\cal C}_{\rm h}<0$. This is a mechanism analogue to the one mentioned for bath-driven dissipation in the main text relying on degenerate transitions (illustrated in Fig. 2).  
Importantly, let us consider now the same above double transition but instead of having the final degenerate states $|n,i\ket$ and $|n,i'\ket$ we take two non-degenerate states $|n,i\ket$ and $|n',i'\ket$. The energy conservation implies that the second final state of $B$ has to be changed from $|\psi_\nu\ket$ to $|\psi_{\nu'}\ket$ such that $E_{\nu'} = e_m-e_{n'} + E_{\mu}$. Consequently, the coherent superposition generated by $U$ disappears after tracing out $B$ (since $\bra \psi_{\nu'}|\psi_\nu\ket=0$). Thus, interestingly, the energy conservation intrinsically prohibits the generation of vertical coherences whereas the generation of horizontal coherences is allowed. \\

\section{Conservation laws}\label{appclaw}
The first step to show the conservation laws is the commutation of the global unitary evolution $U$ with the operation of global block-diagonalisation. We denote by $\Delta_{H_S+H_B}$ such operation. We also denote by $\{\epsilon_k\}_k$ the different energy levels of the ensemble $SB$, and define
\be
\Pi_k:=\sum_{m,\mu; e_m+E_\mu=\epsilon_k} \pi^S_m\pi^B_\mu,
\ee
the projector onto the eigenspace of energy $\epsilon_k$, where $\pi^S_m$ ($\pi^B_\mu$) is itself the projector onto the eigenspace of energy $e_m$ of $S$ ($E_\mu$ of $B$).
Then, the global block-diagonalising operation can be expressed as
\bea
\rho_{SB, t|_{\rm BD}} &=&\Delta_{H_S+H_B} \rho_{SB,t}\nn\\
&=& \sum_k \Pi_k \rho_{SB,t}\Pi_k.
\eea
Importantly, note that $\Delta_{H_S+H_B} \ne \Delta_{H_S}\Delta_{H_B}$, where $\Delta_{H_S}$ denotes the local block-diagonalising operations, $\Delta_{H_S}\rho_{S,t}=\sum_{m} \pi_m^S\rho_{S,t}\pi_m^S$ (and similarly for $B$). From these considerations, one can conclude that $U$ commute with $\Delta_{H_S+H_B}$ if and only if $U$ commute with $\Pi_k$, for all $k$. Due to energy conservation, we indeed have $[U,\Pi_k]=0$, whereas in general $[U,\pi^S_m]\ne0$ and $[U,\pi^B_\mu]\ne0$. Therefore, we obtain that $U$ commute with $\Delta_{H_S+H_B}$ while in general this is not true for $\Delta_{H_S}$ and $\Delta_{H_B}$. This leads to the following important identity,
\be
S(\rho_{SB,t_f|_{\rm BD}}) = S(\rho_{SB,t_i|_{\rm BD}}),
\ee
(where $S(\rho):=-{\rm Tr}\rho\ln\rho$ denotes the von Neumann entropy). This equality implies in particular the global conservation of vertical coherences,
\be
\Delta {\cal C}_{\rm v}^{SB} =0,
\ee
as announced in Eq. (34) of the main text.
Furthermore, thanks to the initial separability of $S$ and $B$, we have,
\bea\label{appsep}
&&\rho_{SB,t_i|_{\rm BD}} = \sum_k \Pi_k \rho_{S,t_i}\rho_{B,t_i}\Pi_k\nn\\
&&= \sum_k~ \sum_{m,\mu; e_m+E_\mu=\epsilon_k}~ \sum_{m',\mu'; e_{m'}+E_{\mu'}=\epsilon_k} \pi_m^S\rho_{S,t_i}\pi_{m'}^S\nn\\
&&\hspace{6cm}\times\pi_\mu^B\rho_{B,t_i}\pi_{\mu'}^B\nn\\
&&= \sum_k~ \sum_{m,\mu; e_m+E_\mu=\epsilon_k} \pi_m^S\rho_{S,t_i}\pi_{m}^S\pi_\mu^B\rho_{B,t_i}\pi_{\mu}^B\nn\\
&&= \sum_m \sum_{\mu} \pi_m^S\rho_{S,t_i}\pi_{m}^S\pi_\mu^B\rho_{B,t_i}\pi_{\mu}^B\nn\\
&&=\rho_{S,t_i|_{\rm BD}}\rho_{B,t_i|_{\rm BD}}(=\rho_{S,t_i|_{\rm BD}}\rho_{B,t_i}),
\eea
where we used $\pi_\mu^B\rho_{B,t_i}\pi_{\mu'}^B= \delta_{\mu,\mu'} \pi_\mu^B\rho_{B,t_i}\pi_{\mu}^B$ since we assumed that $\rho_{B,t_i}$ is a stationary state, $[H_B,\rho_{B,t_i}]=0$. Without this condition, $\rho_{SB,t_i|_{\rm BD}}\ne\rho_{S,t_i|_{\rm BD}}\rho_{B,t_i|_{\rm BD}}$. In other words, when any of the two system $S$ and $B$ is in a stationary state, the global and local block-diagonalising operations are identical, $\Delta_{H_S+H_B}=\Delta_{H_S}\Delta_{H_B}$. This leads to
\be\label{centraleq}
S(\rho_{SB,t_f|_{\rm BD}}) = S(\rho_{SB,t_i|_{\rm BD}}) = S(\rho_{S,t_i|_{\rm BD}})+S(\rho_{B,t_i|_{\rm BD}}). 
\ee
Note that one can prove similarly the following identity, ${\rm Tr}_B \Delta_{H_S+H_B}\rho_{SB} = \Delta_{H_S}\rho_S$, valid for any state  $\rho_{SB}$. Together with \eqref{appsep} and the commutation of $U$ and $\Delta_{H_S+H_B}$ one can formally prove \eqref{appbdid} (corresponding to Eq. (29) of the main text).

The above identity \eqref{centraleq} can be used to refine the conservation law of vertical coherences.
The final relative entropy of vertical coherences of $SB$ can be rewritten as
\bea\label{apppreliminar}
{\cal C}^{SB}_{\rm v}(t_f)&=&S(\rho_{SB,t_f|_{\rm BD}}) - S(\rho_{SB,t_f})\nn\\
&=& S(\rho_{S,t_i|_{\rm BD}})+S(\rho_{B,t_i|_{\rm BD}})- S(\rho_{S,t_i})-S(\rho_{B,t_i})\nn\\
&=&{\cal C}^{S}_{\rm v}(t_i) + {\cal C}^{B}_{\rm v}(t_i)\nn\\
&=&{\cal C}^{S}_{\rm v}(t_i),
\eea
where we used \eqref{centraleq}, the initial separability of $S$ and $B$, and the stationarity of the initial state of $B$, which implies ${\cal C}^{B}_{\rm v}(t_i)=0$. Note that this result was already derived in \cite{Santos_2019}. It means in particular that all the vertical coherences present initially in $\rho_{S,t_i}$ end up in $SR$. This can be made even more precise: the consumption of vertical coherences in $S$ and in $B$ is equal to the final {\it correlated vertical coherences},
\be\label{appconservationen}
-\Delta {\cal C}^S_{\rm v} -\Delta {\cal C}^B_{\rm v} ={\cal C}^{SB}_{\rm c, v}(t_f),
\ee
where 
\be\label{appcoren}
{\cal C}^{SB}_{\rm c, v}(t_f):={\cal C}^{SB}_{\rm v}(t_f) -{\cal C}^{S}_{\rm v}(t_f)-{\cal C}^{B}_{\rm v}(t_f)\geq0
\ee
 quantifies the portion of vertical coherences contained in $SB$ due to correlations between $S$ and $B$ \cite{Tan_2016}. 
Eq. \eqref{appconservationen} is obtained by subtracting ${\cal C}_{\rm v}^S(t_f)$ and ${\cal C}_{\rm v}^B(t_f)$ on both sides of Eq. \eqref{apppreliminar}. 

Still based on \eqref{centraleq}, a similar conservation law can be obtained for horizontal coherences when adding contributions from population convergence. The relative entropy of horizontal coherences ${\cal C}_{\rm h}^{SB}(t_f)$ and the measure of distance ${\cal D}_{\rm th}^{SB}(t_f)$ to a global thermal equilibrium state $\rho_{SB}^{\rm th}(\beta_B)$ are defined in the same way as for $S$. 
Note that due to the energy conservation, any global thermal state is a steady state of $SB$ (valid independently of the initial state of $B$).
 Then, we have,
\bea\label{appconservationnen}
&&{\cal C}_{\rm h}^{SB}(t_f) + {\cal D}_{\rm th}^{SB}(t_f) \nn\\
&&\hspace{0.5cm}= {\rm Tr} \rho_{SB,t_f}[\ln \rho_{SB,t_f|_{\rm BD}} - \ln\rho_{SB}^{\rm th}(\beta_B)]\nn\\
&&\hspace{0.5cm}= -S( \rho_{SB,t_f|_{\rm BD}}\!) \!-\!  {\rm Tr} \rho_{SB,t_f}\!\ln\rho_{SB}^{\rm th}(\beta_B)\nn\\
&&\hspace{0.5cm}= -S( \rho_{S,t_i|_{\rm BD}}) -S( \rho_{B,t_i|_{\rm BD}}) \nn\\
&&\hspace{0.8cm}-  {\rm Tr} \rho_{SB,t_i}\ln\rho_{SB}^{\rm th}(\beta_B)\nn\\
&&\hspace{0.5cm}= -S( \rho_{S,t_i|_{\rm BD}}) -S( \rho_{B,t_i|_{\rm BD}}) \nn\\
&&\hspace{0.8cm}-  {\rm Tr} \rho_{S,t_i}\ln\rho_{S}^{\rm th}(\beta_B)-  {\rm Tr} \rho_{B,t_i}\ln\rho_{B}^{\rm th}(\beta_B)\nn\\
&&\hspace{0.5cm}= {\cal C}_{\rm h}^{S}(t_i) + {\cal D}_{\rm th}^{S}(t_i) +{\cal C}_{\rm h}^{B}(t_i) + {\cal D}_{\rm th}^{B}(t_i).
\eea
Since $S$ and $B$ are initially uncorrelated, the identity \eqref{appconservationnen} implies the following conservation law
\be\label{appglobalconservation}
\Delta {\cal C}_{\rm h}^{SB}+ \Delta {\cal D}_{\rm th}^{SB}=0.
\ee
Defining concepts of correlated horizontal coherences,
\be 
{\cal C}^{SB}_{\rm c, h}(t):={\cal C}^{SB}_{\rm h}(t) -{\cal C}^{S}_{\rm h}(t)-{\cal C}^{B}_{\rm h}(t_f)\geq0,
\ee
and correlated population distance to the thermal equilibrium state,
\be 
{\cal D}^{SB}_{\rm c, th}(t):={\cal D}^{SB}_{\rm th}(t) -{\cal D}^{S}_{\rm th}(t)-{\cal D}^{B}_{\rm th}(t)\geq0,
\ee
in a similar way as \eqref{appcoren}, one can obtain a refined statement in the same form as \eqref{appconservationen},
\bea\label{appglobalconservation2}
-\Delta {\cal C}^S_{\rm h} \!-\!\Delta {\cal C}^B_{\rm h}\!-\!\Delta {\cal D}^S_{\rm th} \!-\!\Delta {\cal D}^B_{\rm th} \!={\cal C}^{SB}_{\rm c, h}(t_f)+{\cal D}^{SB}_{\rm c, th}(t_f).\nn\\
\eea
The above identity means that the consumption of horizontal coherences in $S$ and $B$ plus the population convergence to the local equilibrium state is equal to the final correlated horizontal coherences plus correlated distance. As a corollary, horizontal coherences {\it are not conserved}, as expected. 
 Quite curiously, \eqref{appglobalconservation} and \eqref{appglobalconservation2} can be established using any 
 global thermal state as steady state. Thus, even though $\rho_{S}^{\rm th}(\beta_B)={\rm Tr}_B\rho_{SB}^{\rm th}(\beta_B)$ might not be an equilibrium state of $S$ for $\Lambda$ (since $\Lambda$ and its equilibrium states depend on the initial state of $B$), still, the conservation laws hold.

Considering a situation analogue to Section \ref{secresource} of the main text, 
 one can explicitly obtain a situation with $-\Delta {\cal D}_{\rm th}^S <0$. 
More precisely, one can take $B$ initially in a thermal state at inverse temperature $\beta_B$ and $S$ initially in a state $\rho_{S,0}=\rho_S^{\rm th}(\beta_B)+\chi$ composed of populations thermally distributed at the same inverse temperature $\beta_B$ and $\chi$ containing horizontal coherences (remembering that vertical coherences have no effect here). Defining the entropy production from the thermal equilibrium state $\rho_S^{\rm th}(\beta_B)$ one has initially ${\cal D}_{\rm th}^S(0)=0$ whereas ${\cal D}_{\rm th}^S(t=+\infty)>0$ since $S$ gains (loses) energy if the horizontal coherences contained in $\chi$ are such that $c^{+}>c^{-}$ ($c^{+}<c^{-}$), see Eq. (18) of the main text. Consequently, $-\Delta {\cal D}_{\rm th}^S <0$.  
Using the above conservation law \eqref{appglobalconservation} (corresponding to Eq. (37) of the main text) one can also see that the presence of initial horizontal coherences can lead to $-\Delta {\cal D}_{\rm th}^{SB}<0$, corresponding to a global divergence of the populations from the thermal equilibrium distribution. 

Finally, using \eqref{appglobalconservation} we can show explicitly that the generation of horizontal coherences affect the energy exchanges. This can be seen as follows. Considering $B$ in a thermal state a inverse temperature $\beta_B$ and defining the population convergence with respect to the global thermal state at inverse temperature $\beta_B$, we have the following identity, $-\Delta {\cal D}_{\rm th}^{SB} = -\beta_B {\cal F}_D^{SB}$. From the energy conservation and using \eqref{appglobalconservation}, one can rewrite the final diagonal entropy as 
\be
S(\rho_{SB,t_f|{\rm D}}) = S(\rho_{SB,t_i|{\rm D}}) + \Delta {\cal C}_{\rm h}^{SB}.
\ee
Then, one can see that the final global diagonal entropy is strictly increased in a scenario where horizontal coherences are generated when compared to a situation where no horizontal coherences is generated. It means that the final populations (of $S$ or $B$, or both) are necessarily altered by the generation of horizontal coherences, implying that both the final energy and the energy exchange are altered. Thus, through the conservation laws, one recovers the observation made in \cite{bathinducedcoh,bathinducedcohTLS}: bath-induced coherences affects the energy exchanges. \\

 \section{Effects in thermal machines performances}\label{appeffecttm}
 In this paragraph we look at a cyclic thermal machine with a working medium $S$ containing degenerate energy levels. The working medium $S$ is successively in contact with a cold bath at temperature $T_c$ and a hot bath at temperature $T_h$. Due to the degeneracy of $S$, the coupling with each bath might involve degenerate transitions (see Fig. 2 of the main text), resulting in coupled dynamics of the horizontal coherences and populations. As seen in the main text, this might lead to negative contributions to the entropy production. Conversely, one can imagine a situation where the coupling with the baths involves no degenerate transitions (and therefore no negative contribution to the entropy production). As a result, the two kinds of dynamics might result in different entropy production (as for instance in the illustration in Section \ref{secnegcontrib} of the main text). The aim of this paragraph is to compare the performances of the ``coherent'' thermal machine (when horizontal coherences and populations are coupled) to the performances of the ``incoherent'' thermal machine (when horizontal coherences and populations are not coupled). 
    To simplify the discussion we consider a simple Otto cycle \cite{Scully_2002, Quan_2007} composed of the usual succession of one adiabatic stroke, one isochoric stroke in contact with the cold bath, one adiabatic stroke (which can be the reverse of the first one), and finally one isochoric stroke in contact with the hot bath. For the incoherent machine we denote by $Q_c$, $Q_h$, and $\Sigma$ the heat exchanged per cycle with the cold bath (i.e. during the isochore with the cold bath), the hot bath and the entropy production per cycle, respectively. For the coherent one we denote the corresponding quantities by $Q_c^*$, $Q_h^*$, and $\Sigma^*$. During one full cycle, the entropy variation $\Delta S$ of $S$ is null and the second law can be expressed as 
    \be
   0= \Delta S = \Sigma +\frac{Q_c}{T_c} + \frac{Q_h}{T_h} .
    \ee
   Similarly, for the coherent machine we have $ 0=\Delta S = \Sigma^* +\frac{Q_c^*}{T_c} + \frac{Q_h^*}{T_h} = 0$. Then, assuming $\Sigma \ne \Sigma^*$ (as a result of negative contributions to the entropy production) and defining $\Delta Q_c := Q^*_c-Q_c$ and $\Delta Q_h := Q^*_h-Q_h$, we have necessarily $\Delta Q_c\Delta Q_h \ne 0$. Considering the work extraction operating mode of the machine, we denote by $W = -Q_c-Q_h\leq 0$ the work extracted per cycle, and by $\eta = \frac{|W|}{Q_h}$ the associated efficiency (for the incoherent machine). For the coherent machine the corresponding quantities are denoted by $W^*$ and $\eta^*$. 
   
   Even though $\Delta Q_c\Delta Q_h \ne0$, we might have $W=W^*$. If so, one can show that the efficiency are necessarily different. More precisely, some simple manipulations give 
   \be
   \eta^* = \eta + \Delta Q_h \frac{W}{Q_hQ_h^*}.
   \ee
   In particular, still assuming work extraction, we have $\eta^*>\eta$ if and only if $\Sigma^*<\Sigma$. 
   
   Conversely, even though $\Delta Q_c\Delta Q_h \ne0$, we might have $\eta=\eta^*$. Similarly, we can show that such situation implies 
   \be
   |W^*| = |W| + \left(1+ \frac{Q_c}{Q_h}\right)\Delta Q_h. 
   \ee
   In particular, one has $|W^*|>|W|$ if and only if $\Sigma^*>\Sigma$.  
   
   Then, we can draw the conclusion that any change in the entropy production per cycle inevitably affects (positively or negatively) the power or efficiency (or both) of the machine. Therefore, the alteration of the entropy production described throughout the paper can have important implication for thermal machines. As an illustration, taking for $S$ an ensemble of spins one can show that $\eta^*=\eta$ always holds and that $\Sigma^*>\Sigma$ for adequately chosen values of $T_c$ and $T_h$. This  implies that $|W^*|>|W|$, recovering the results of \cite{bathinducedcoh}. For bad choice of $T_c$ and $T_h$, $\Sigma^*<\Sigma$ and the extracted work per cycle is degraded.

 Note that this analysis fails for vertical coherences because they break the cycle: if vertical coherences are introduced at the beginning of the cycle, they are not recovered at the end of the cycle (at least when considering adiabatic strokes). By contrast, horizontal coherences do not need to be introduced, they are induced by the bath through collective coupling (or degenerate transitions). 
 Considering more complex cycles by introducing non-adiabatic strokes, vertical coherences (in the eigenbasis of the instantaneous Hamiltonian of the working medium) can be generated by the external driving \cite{Camati_2019}, so that one has the possibility to cyclically recover vertical coherences, and a similar analysis as the above one may apply.


%
%
%
%
%
%



\begin{thebibliography}{1} 
\bibitem{Spohn_1978} H. Spohn, Journal of Mathematical Physics {\bf 19}, 1227 (1978).
\bibitem{Spohn_1978b} H. Spohn and J. L. Lebowitz, Adv. Phys. Chem. {\bf 38}, 109-142 (1978).
\bibitem{Alicki_1979} R. Alicki, J. Phys. A: Math. Gen. {\bf 12}, L103 (1979).
\bibitem{Parrondo_2009} J. M. R. Parrondo, C. Van den Broeck and R. Kawai, New J. Phys. {\bf 11}, 073008 (2009).
\bibitem{Deffner_2011} S. Deffner and E. Lutz, Phys. Rev. Lett. {\bf 107}, 140404 (2011).
\bibitem{Santos_2017} J. P. Santos, G. T. Landi, and M. Paternostro, Phys. Rev. Lett. {\bf 118}, 220601 (2017).
\bibitem{Brunelli_2018} M. Brunelli, L. Fusco, R. Landig, W. Wieczorek, J. Hoelscher-Obermaier, G. Landi, F. L. Semi{\~a}o, A. Ferraro, N. Kiesel, T. Donner, G. De Chiara, and M. Paternostro, Phys. Rev. Lett. {\bf 121}, 160604 (2018).
\bibitem{Santos_2019} J. P. Santos, L. C. C{\'e}leri, G. T. Landi and M. Paternostro, npj Quantum Information {\bf 5}:23 (2019).
\bibitem{Batalhao_2018} T.B. Batalh{\~a}o, S. Gherardini, J.P. Santos, G.T. Landi, M. Paternostro, (2018) {\it Characterizing Irreversibility in Open Quantum Systems}. In: Binder F., Correa L., Gogolin C., Anders J., Adesso G. (eds) {\it Thermodynamics in the Quantum Regime}. Fundamental Theories of Physics, vol 195. Springer, Cham.
\bibitem{Li_2019} Sheng-Wen Li, Entropy {\bf 21}, 111 (2019).
\bibitem{Barato_2015} A. C. Barato and U. Seifert, Phys. Rev. Lett. {\bf 114}, 158101 (2015)
\bibitem{Gingrich_2016} T. R. Gingrich, J. M. Horowitz, N. Perunov, and J. L. England, Phys. Rev. Lett. {\bf 116}, 120601 (2016).
\bibitem{Pietzonka_2016} P. Pietzonka, A. C. Barato, and U. Seifert, Phys. Rev. E {\bf 93}, 052145 (2016).
\bibitem{Pietzonka_2018} P. Pietzonka and U. Seifert, Phys. Rev. Lett. {\bf 120}, 190602 (2018).
\bibitem{Guarnieri_2019} G. Guarnieri, G. T. Landi, S. R. Clark, and J. Goold, arXiv:1901.10428.
\bibitem{Timpanaro_2019} A. M. Timpanaro, G. Guarnieri, J. Goold, and G. T. Landi, Phys. Rev. Lett. {\bf 123}, 090604 (2019).
\bibitem{Su_2019} S. Su, W. Shen, J. Du, and J. Chen, arXiv:1904.04113.
\bibitem{Holubec_2018} V. Holubec and A. Ryabov, Phys. Rev. Lett. {\bf 121}, 120601 (2018).



\bibitem{Francica_2019} G. Francica, J. Goold, and F. Plastina, Phys. Rev. E {\bf 99}, 042105 (2019).
\bibitem{Camati_2019} P. A. Camati, J. F. G. Santos, and R. M. Serra, Phys. Rev. A {\bf 99}, 062103 (2019).
\bibitem{bathinducedcoh} C. L. Latune, I. Sinayskiy, and F. Petruccione, Phys. Rev. Research {\bf1}, 033192 (2019).

\bibitem{Cuetara_2016} G. B. Cuetara, M. Esposito, and G. Schaller, Entropy {\bf 18}, 447 (2016).
\bibitem{Savchenko_1998} V. I. Savchenko, N. J. Fisch, A. A. Panteleev, and A. N. Starostin, Phys. Rev. A {bf 59}, 708 (1998).
\bibitem{Agarwal_2001} G. S. Agarwal, and Sunish Menon, Phys. Rev. A {\bf 63}, 023818 (2001).
\bibitem{Dag_2016} C. B. Da{\u g}, W. Niedenzu, {\"O}. E. M\"{u}stecapl\i o\v{g}lu, and G. Kurizki, Entropy {\bf 18}, 244 (2016).

\bibitem{paperapptemp} C. L. Latune, I. Sinayskiy, and F. Petruccione,  Quantum Sci. Technol. {\bf 4}, 025005 (2019).

\bibitem{Cakmak_2017} B. {\c C}akmak, A. Manatuly, and {\"O}. E. M\"{u}stecapl\i o\v{g}lu, Phys. Rev. A {\bf 96}, 032117 (2017).
\bibitem{bathinducedcohTLS} C. L. Latune, I. Sinayskiy, and F. Petruccione, Phys. Rev. A {\bf 99}, 052105 (2019).
\bibitem{Dag_2019} C. B. Da{\u g}, W. Niedenzu, F. Ozaydin, {\"O}. E. M\"{u}stecapl\i o\v{g}lu, and G. Kurizki, J. Phys. Chem. C {\bf 123}, 7, 4035-4043 (2019).


%

\bibitem{heatflowreversal} C. L. Latune, I. Sinayskiy, and F. Petruccione, Phys. Rev. Research {\bf1}, 033097 (2019).
\bibitem{Skribanowitz_1973} N. Skribanowitz, l. P. Herman, J. C. MacGillivray, and M. S. Feld, Phys. Rev. Lett. {\bf 30}, 309 (1973).
\bibitem{Gross_1976} M. Gross, C. Fabre, P. Pillet, and S. Haroche, Phys. Rev. Lett. {\bf 36}, 1035 (1976).
\bibitem{Raimond_1982} J. M. Raimond, P. Goy, M. Gross, C. Fabre, and S.
Haroche, Phys. Rev. Lett. {\bf 49}, 117 (1982).
\bibitem{Devoe_1996} R. G. DeVoe and R. G. Brewer, Phys. Rev. Lett. {\bf 76}, 2049 (1996).
\bibitem{Gross_1982} M. Gross and S. Haroche, Physics Reports  {\bf 93}, 301-396 (1982).
\bibitem{Kim_2002} M. S. Kim, J. Lee, D. Ahn, and P. L. Knight, Phys. Rev. A {\bf 65}, 040101 (2002).
\bibitem{Plenio_2002} M. B. Plenio and S. F. Huelga, Phys. Rev. Lett. {\bf 88}, 197901 (2002).
\bibitem{Benatti_2003} F. Benatti, R. Floreanini, and M. Piani, Phys. Rev. Lett. {\bf 91}, 070402 (2003). 
\bibitem{Muschik_2011} C. A. Muschik, E. S. Polzik, and J. I. Cirac, Phys. Rev. A {\bf 83}, 052312 (2011).
\bibitem{Krauter_2011} H. Krauter, C. A. Muschik, K. Jensen, W. Wasilewski, J. M. Petersen,
J. I. Cirac, and E. S. Polzik, Phys. Rev. Lett. {\bf 107} 080503 (2011).
\bibitem{Cotlet_2014} O. Cotlet and B. W. Lovett, New J. Phys. {\bf 16}, 103016 (2014).




\bibitem{Janzing_2000} D. Janzing, P. Wocjan, R. Zeier, R. Geiss, and T. Beth, Int. J.
Theor. Phys. {\bf 39}, 2717 (2000).
\bibitem{Petruccione_Book} H. Breuer and F. Petruccione, {\it Theory of Open Quantum Systems}, (Oxford, Oxford, 2002).
\bibitem{Jagadish_2018} V. Jagadish and F. Petruccione, Quanta {\bf 7}, 54-67 (2018).
\bibitem{Kosloff_2013} R. Kosloff, Entropy {\bf 15}, 2100-2128 (2013).
\bibitem{Kosloff_2014} R. Kosloff and A. Levi, Annu. Rev. Phys. Chem. {\bf 65}:365–93 (2014).
\bibitem{Lindblad_1975}  G. Lindblad, Commun. Math. Phys. {\bf 40}, 147 (1975).
\bibitem{Baumgratz_2014} T. Baumgratz, M. Cramer, and M. B. Plenio, Phys. Rev. Lett. {\bf 113}, 140401 (2014).
\bibitem{Manzano_2019} G. Manzano, R. Silva, and J. M. R. Parrondo, Phys. Rev. E {\bf 99}, 042135 (2019).
\bibitem{Marvian_2016a} I. Marvian, R. W. Spekkens, and P. Zanardi, Phys. Rev. A {\bf93}, 052331 (2016).
\bibitem{Marvian_2016b} I. Marvian and R. W. Spekkens, Phys. Rev. A {\bf94}, 052324 (2016).
\bibitem{Nielsen_Book} M. A. Nielsen and I. L. Chuang, {\it Quantum Computation and Quantum Information} (10th Anniversary Edition, Cambridge University Press, 2010).
\bibitem{Alicki_2014} R. Alicki, arXiv:1401.7865.
\bibitem{Alicki_2015} R. Alicki and D. Gelbwaser-Klimovsky, New J. Phys. {\bf 17}, 115012 (2015).
\bibitem{Cohen_Book} C. Cohen-Tannoudji, J. Dupont-Roc, and G. Grynberg, {\it Processus d'interaction entre photons et atomes}, (EDP Science/CNRS {\'E}ditions, Paris, 2001).
\bibitem{autonomousmachines} C. L. Latune, I. Sinayskiy, and F. Petruccione, Sci. Rep. {\bf 9}:3191 (2019).

\bibitem{Patnaik_1999} A. K. Patnaik and G. S. Agarwal, Phys. Rev. A {\bf 59}, 3015 (1999).
\bibitem{Koslov_2006} V. V. Kozlov, Y. Rostovtsev, and M. O. Scully, Phys. Rev. A {\bf 74}, 063829 (2006).
\bibitem{Tscherbul_2014} T. V. Tscherbul and P. Brumer, Phys. Rev. Lett. {\bf 113}, 113605 (2014).
\bibitem{Dodin_2016} A. Dodin, T. V. Tscherbul, and P. Brumer, The Journal of Chemical Physics {\bf 144}, 244108 (2016).



\bibitem{Partovi_2008} M. H. Partovi, Phys. Rev. E {\bf77}, 021110 (2008).
\bibitem{Jennings_2010} D. Jennings and T. Rudolph, Phys. Rev. E {\bf81}, 061130 (2010).
\bibitem{Micadei_2019} K. Micadei, J. P. S. Peterson, A. M. Souza, R. S. Sarthour, I. S. Oliveira, G. T. Landi, T. B. Batalhão, R. M. Serra, and E. Lutz, Nat. Commun. {\bf10}, 2456 (2019).
\bibitem{Henao_2018} I. Henao, and R. M. Serra, Phys. Rev. E {\bf 97}, 062105 (2018).



\bibitem{Horodecki_2013} M. Horodecki1, and J. Oppenheim, Nat. Commun. {\bf4}, 2059 (2013).
\bibitem{Meer_2017} R. v. d. Meer, N. H. Y. Ng, and S. Wehner, Phys. Rev. A {\bf96}, 062135 (2017).
\bibitem{Muller_2018} Markus P. M{\"u}ller, Phys. Rev. X {\bf8}, 041051 (2018).
\bibitem{Perry_2018} C. Perry, P. {\'C}wikli{\'n}ski, J. Anders, M. Horodecki, and J. Oppenheim, Phys. Rev. X {\bf 8}, 041049 (2018).



\bibitem{Tan_2016} K. C. Tan, H. Kwon, C.-Y. Park, and H. Jeong, Phys. Rev. A {\bf 94}, 022329 (2016). 
%
%
\bibitem{Xia_1996} H.-R. Xia, C.-Y. Ye, and S.-Y. Zhu, Phys. Rev. Lett. {\bf 77}, 1032 (1996).
\bibitem{Norcia_2016} M. A. Norcia, M. N. Winchester, J. R. K. Cline, J. K. Thompson, Sci. Adv. {\bf2}, e1601231 (2016).
\bibitem{Norcia_2018} M. A. Norcia, R. J. Lewis-Swan, J. R. K. Cline, B. Zhu, A. M. Rey, and J. K. Thompson, Science {\bf361}, 259 (2018).
\bibitem{Barberena_2019} D. Barberena, R. J. Lewis-Swan, J. K. Thompson, and A. M. Rey, Phys. Rev. A {\bf 99}, 053411 (2019).

\bibitem{Smirne} A. Smirne, D. Egloff, M. G. D{\'i}az, M. B. Plenio and S. F. Huelga, Quantum Sci. Technol. {\bf4}, 01LT01 (2019).
\bibitem{Lostaglio_2015} M. Lostaglio, K. Korzekwa, D. Jennings, and T. Rudolph, Phys. Rev. X {\bf 5}, 021001 (2015).
\bibitem{Streltsov_2017} A. Streltsov, G. Adesso, M. B. Plenio, Rev. Mod. Phys. {\bf 89}, 041003 (2017).
\bibitem{Llobet_2019} M. Perarnau-Llobet and R. Uzdin, New J. Phys. {\bf 21}, 083023 (2019).
\bibitem{Miller_2019} H. J. D. Miller, M. Scandi, J. Anders, and M. Perarnau-Llobet, arxiv:1905.07328.


\bibitem{Tscherbul_2015} T. V. Tscherbul and P. Brumer, The Journal of Chemical Physics {\bf142}, 104107 (2015).

\bibitem{Zhou_2001} P. Zhou, Phys. Rev. A {\bf 63}, 023810 (2001).
\bibitem{Wood_2014} C. J. Wood, T. W. Borneman, and D. G. Cory, Phys. Rev. Lett. {\bf 112}, 050501 (2014).
\bibitem{Wood_2016}  C. J. Wood and D. G. Cory, Phys. Rev. A {\bf 93}, 023414 (2016).
\bibitem{Hama_2018} Y. Hama, W. J. Munro, and K. Nemoto, Phys. Rev. Lett. {\bf 120}, 060403 (2018).
\bibitem{Niedenzu_2018} W. Niedenzu and G. Kurizki, New J. Phys. {\bf 20}, 113038 (2018).
\bibitem{LandauLifshitz_Book} L. D. Landau and E. M. Lifshitz, {\it Quantum Mechanics: Non-Relativistic Theory}, (Pergamon Press, Oxford) 1977.
\bibitem{Sakurai_Book} J. J. Sakurai, {\it Modern Quantum Mechanics} (Addison Wesley, Reading, MA, 1993).
\bibitem{Scully_2002} M. O. Scully, Phys. Rev. Lett. {\bf 88}, 050602 (2002).
\bibitem{Quan_2007} H. T. Quan, Y.-x. Liu, C. P. Sun, and F. Nori, Phys. Rev. E {\bf 76}, 031105 (2007).





\end{thebibliography}
\end{document}